\documentclass[preprint,showpacs,preprintnumbers,amsmath,amssymb]{revtex4}

\usepackage{graphicx}
\usepackage{dcolumn}
\usepackage{bm}

\newcommand{\be}{\begin{equation}}  
\newcommand{\ee}{\end{equation}}  
\newcommand{\bear}{\begin{eqnarray}}  
\newcommand{\eear}{\end{eqnarray}}  
\newcommand{\ba}{\begin{array}}  
\newcommand{\ea}{\end{array}}

\newskip\humongous \humongous=0pt plus 1000pt minus 1000pt

\newif\ifdtup

\def\oldreffmt#1{\rlap{[#1]} \hbox to 2\parindent{}}

\def\figfmt#1{\rlap{Figure {#1}} \hbox to 1in{}}  
  
\def\ie{\hbox{\it i.e.}{}}	\def\etc{\hbox{\it etc.}{}}  
\def\eg{\hbox{\it e.g.}{}}

\def\tr{\mathop{\rm tr}}  
\def\Tr{\mathop{\rm Tr}}

\def\bra#1{\left\langle #1\right|}  
\def\ket#1{\left| #1\right\rangle}

\def\slash#1{#1\!\!\!/\!\,\,}  
\def\beq{\begin{equation}}  
\def\eeq{\end{equation}}  
\def\bea{\begin{eqnarray}}  
\def\eea{\end{eqnarray}}  
\def\half{\frac{1}{2}}  
  
\def\bq{\begin{quote}}  
\def\eq{\end{quote}}

\def\half{\frac{1}{2}}       

\relax  

\newdimen\tdim  
\tdim=\unitlength  
\def\bar{\overline}

\begin{document}

\preprint{FERMILAB-Pub-06/010-T}


\title{
Anomalies, Chern-Simons Terms\\
and Chiral Delocalization in Extra Dimensions }

\author{Christopher T. Hill}

\email{hill@fnal.gov, }

\affiliation{
 {{Fermi National Accelerator Laboratory}}\\
{{\it P.O. Box 500, Batavia, Illinois 60510, USA}}
}%

\date{January 18, 2006}

\begin{abstract}
Gauge invariant topological interactions, such as the 
$D=5$ Chern-Simons term, are required in models in
extra dimensions that split anomaly free representations.
The Chern-Simons term is necessary to maintain the overall
anomaly cancellations of the theory, but it
can have significant, observable, physical effects. 
The CS-term locks the KK-mode parity to the
parity of space-time, leaving a single parity symmetry.
It leads to new processes amongst KK-modes
\eg, the decay of a KK-mode
to a 2-body final state of KK-modes. 
A formalism for the effective interaction 
amongst KK-modes is constructed, and the decay of a KK-mode
to KK-mode plus zero mode is analyzed as an example. We elaborate
the general KK-mode current and anomaly structure of these theories.  
This includes a detailed study of the triangle diagrams
and the associated ``consistent anomalies'' for Weyl 
spinors on the boundary branes.
We also develop the non-abelian formalism. We illustrate
this by showing in a simple way how a $D=5$ Yang-Mills 
``quark flavor'' symmetry leads to the $D=4$ chiral lagrangian of mesons
and the quantized Wess-Zumino-Witten term.

\end{abstract}

\maketitle

\section{\bf Introduction}

We explore the physics of a
Chern-Simons  term in a model of  a gauge theory 
in a compactified extra dimension.
A Chern-Simons term 
leads to new physical processes, \ie, new interactions
amongst the KK-modes of gauge fields,
characterized by a quantized coefficient.
The Chern-Simons term locks KK-mode parity
to the parity of space-time via the 5-$\epsilon$ symbol
permitting processes in which KK-parity is violated,
while KK-parity combined with $D=4$ space-time parity is conserved.
That is, KK-mode parity is a spurious symmetry and is independent 
of space-time parity, until a Chern-Simons term appears, which
unites these into a single parity symmetry.

The effects of Chern-Simons terms are intertwined with the anomalies
of matter fields. In the present discussion
matter fields will be localized on boundary branes.
Since the Chern-Simons term is a bulk interaction, it probes
physics away from the boundary (branes). 
Taken together, however, the anomalies and Chern-Simons
terms produce  gauge invariant amplitudes containing
new physics.  Perhaps the most novel new physics is
the decay of
a single KK-mode into a two KK-mode final state. We study
the latter process in this paper as an explicit example of
the formalism.

It should not be surprising that
there is nontrivial physics associated with the CS term.
For example, in  the case of $D=3$ QED, the CS term 
gives the photon a mass \cite{schonfeld}. 
Moreover, the $D=5$ Yang-Mills CS term for $SU(N)$ can be 
deformed into a Wess-Zumino-Witten (WZW) term \cite{wess,witten},
of an $SU(N)_L\times SU(N)_R$ 
chiral lagrangian of mesons, under certain compactification
schemes \cite{zachos} 
(we give a simple derivation of this in Section V of the present
paper \cite{hill4}).
Many essential physical processes in QCD, such as
$\pi\rightarrow 2\gamma$,
$\phi\rightarrow K\bar{K}$, and $\phi\rightarrow 3\pi$, etc.,
are controlled by the WZW term. In particular,
the pion parity in a chiral lagrangian is defined by the WZW term
via the 4-$\epsilon$ symbol 
and would be a spurious symmetry without it. This is the 
analogue of the fate of KK-parity
in the presence of the CS term in the $D=5$ bulk.

Split fermion theories \cite{nima} are 
compelling in that anomalous representations of chiral fermions occur
in a delocalized way in extra dimensions. Given that
the standard model involves flavor and chirality in a
nontrivial (non-vectorlike) way, 
it would seem plausible that chiral delocalization
would occur if extra dimensions should exist. Chirality is then,
in a sense, emergent from the spatial localization in
extra dimensions. A key discriminant
for such a theory is, indeed, likely to be the KK-mode parity
violation through CS term interactions.

We study in detail 
a $U(1)$ theory, but we also indicate
how things work in non-abelian theories as well. 
Our goal is to develop a solid formalism for correctly
obtaining gauge invariant amplitudes. In the non-abelian case we 
use this formalism to demonstrate how
the Wess-Zumino-Witten term, with its quantized coefficient, arises
by morphing a $D=5$ Yang-Mills theory of quark flavor into
a compactified $D=4$ theory, where it becomes a chiral
lagrangian of mesons, the ``mesons'' being the $A^5$ gauge fields
\cite{zachos}.

In $D=5$, for a $U(1)$ gauge theory, there exists a CS term of the form 
$ \epsilon_{ABCDE}A^A \partial^B A^C  \partial^D A^E $, and we'll consider
this operator to be part of a lagrangian of a 
$D=5$ theory compactified to $D=4$ with boundary
branes. Under a gauge transformation
this term produces anomalies on the boundary branes, which
we'll refer to as the ``{\em CS anomalies}." Anomalies represent
nonconserved currents, \ie,
$\partial_Aj^A =$(anomaly).
Any gauge
theory would be {\em a priori} inconsistent with such anomalies since the
equation of motion, $\partial_A F^{AB} = j^B$, implies
$\partial_A j^A= 0 $, by the antisymmetry of $F^{AB}$.
We thus require some mechanism to cancel anomalies, and
this typically implies something like chiral matter fields attached 
to different boundary branes that produce their own ``{\em matter anomalies}''
\cite{jackiw,adler,bardeen,jackiw2}.
The CS anomalies must then cancel against the matter anomalies \cite{skiba}.

The relevant fermionic anomalies on the boundaries
are the ``consistent anomalies,'' \ie, the form
that arises directly from Feynman diagrams.
We explicitly verify in detail
the results of Bardeen \cite{bardeen} for the consistent anomalies
of massless Weyl spinors, (and the massive ``left-right symmetric'' case)
arising from the effective action operators 
in Appendix C. A key point, often confused in the 
literature, is that
the Weyl spinor anomaly (a consistent anomaly)  {\em is not 
``half the Dirac spinor axial current anomaly of QED,'' } \cite{adler} 
(the ``covariant anomaly'').
The consistent and covariant forms of
the anomalies differ by the addition of a counterterm to the action that
makes the vector (electromagnetic) current conserved, even in the presence of
a background axial vector photon.
Remarkably (though perhaps not surprisingly), we find that this counterterm 
is precisely  ``boundary term''
that arises when we integrate out the fermions on the boundary branes in the
large fermion mass limit.  In combination 
with the CS anomaly, which has the form of
the consistent anomaly, this maintains the zero-mode gauge invariance,
and it shifts the lowest axial vector current anomaly into the covariant form.
We obtain the resulting tower of KK-mode covariant
current anomalies in Appendix B.

Our formal problem, in part, is 
to bring the full 
effective action into a manifestly gauge invariant form. 
Indeed, one of the beautiful aspects of gauge field theories 
under compactification
is the manner in which the KK-mode mass terms are automatically
generated in a manifestly gauge invariant way. The modes $A_\mu^n$  
become packaged together
with their longitudinal components $\partial_\mu A_5^n $, as
they acquire mass -- they become gauge invariant
``Stueckelberg fields.'' 
Once we are assured that the action can be brought into Stueckelberg
combinations of massive fields, 
then we are assured of unambigous Feynman rules that respect 
gauge invariance for the massive fields.
We find that a Wilson line gauge transformation that brings
$A^5 =0$ in the bulk simultaneously packages
all of the massive KK-modes in the CS-term into
Stueckelberg form.  This leads to a remarkably 
compact formula for the CS-term part
of the effective action. In the case
of chiral fermions on the boundaries, the
triangle diagrams maintain the zero-mode gauge invariance. 
A physical amplitude is thus a sum of the CS term contribution
in the bulk plus a boundary term that comes from the chiral
fermions on the branes.
In the case of delocalized chiral fermions with
a Wilson line mass term, and a $U(1)$ gauge group,
the boundary term arises from triangle diagrams involving the
fermions on the branes.

We consider
two distinct ways to implement matter fields.
The  conceptually simpler method employs axions, $\phi_L$ and $\phi_R$, on
the $x^5=0$ ($L$) and $x^5=R$ ($R$)  coupled
to $F\tilde{F}(0)$ and $F\tilde{F}(R)$ on 
their respective boundary branes.
This construction involves
no triangle diagrams,  yet it demonstrates, as a matter of 
principle that the CS term 
bulk interactions are physical. However, this theory
is somewhat less interesting because here
the zero mode photon itself becomes massive, by eating one linear
combination, $\phi_L+\phi_R$, of the axions. A massive Stueckelberg
photon field now appears explicitly in the CS term.
A second, orthogonal, combination $\phi_L-\phi_R$
remains as a physical uneaten axion. 
We discuss this in Appendix A.

We are mainly interested 
in QED in $D=5$ with ``split chiral electrons''
on the boundary branes. Though QED
is a vectorlike theory, if we want to promote
it to $D=5$, yet maintain a naturally small electron mass
(compared to $1/R$, the compactification scale),
then we require chiral delocalization
\eg, split chiral electrons on the boundary branes.
We thus place $\psi_L$ on the left-brane $x^5=0$ and $\psi_R$
on the right-brane $x^5=R$.  The electron mass term is then a
bilocal operator involving a Wilson line, 
$m \bar{\psi}_L\exp(\int i\; dx^5 A_5)\psi_R +h.c.$ where
the integral runs from $x^5=0$ to $x^5=R$.
This is the technically natural setting for QED in $D=5$, and
it necessarily contains the CS-term due to the split
fermion representation. 
The anomalies on the branes cancel the CS term
anomalies for a particular
choice of the coefficient of the CS term. Thus the 
CS term is ``quantized.'' In this model the photon stays
massless, and the 
full effective interaction consisting of
the CS term, together with the 
fermion triangle loop contributions is now explicitly gauge invariant
for all modes \footnote{
Note, of course, that the branes can be  explicitly 
constructed by coupling a $D=5$
bulk electron to a kink plus antikink solition. 
This would yield the brane chiral
electrons as domain wall fermions. Our branes are presently
imposed by hand, yet the present model can be viewed as a description
of domain wall fermions in the limiting case of a thin domain wall,
and a large fermionic Dirac mass in the bulk.}.

The non-abelian theory can be developed along parallel lines.
It can also contain a $\tr(AdAdA) + ...$ CS-term, and boundary brane
chiral fermions. 
We develop the non-abelian case to the point of establishing
the quantized coefficient of the CS-term. We then revisit the 
problem of morphing a $D=5$ Yang-Mills theory of quark flavor into
the low energy Wess-Zumino-Witten term by showing how
the $\Tr(\pi d\pi d\pi d\pi d\pi )$ term arises with the Witten quantization
in a theory with flipped orbifold boundary conditions \cite{zachos}. 

Most technical details have been
relegated to a series of Appendices. The main text attempts
to give the sequential arguments in a more conceptual form. 
We have included relevant background issues concerning anomalies
to enhance the reader's familiarity with some of the subtleties.

\section{The $U(1)$ Case: General Argument}

Consider a $U(1)$ gauge theory 
in $D=5$, with the covariant derivative, field strength and
kinetic term lagrangian density:
\bea
D_A & = & \partial_A - iA_A \; ,
\qquad F_{AB} = i[D_A, D_B]\; ,
\qquad
L_0  =  -\frac{1}{4\widetilde{e}^2} F_{AB}F^{AB}\; .
\eea
We define the gauge fields to have canonical
dimension for $D=4$, \ie, $A\sim M^1$,
where then $1/\widetilde{e^2}$ has dimension of $M^1$
(the e.t.c. is then $[A_i(\vec{x}),\dot{A_j}(\vec{y})] = 
i\widetilde{e}^2\delta^4(\vec{x}-\vec{y})$).

The  theory admits 
a Chern-Simons term, defined by 
the local lagrangian density:
\bea
L_{CS} & = & c \;\epsilon^{ABCDE} A_{A}\partial_B A_C\partial_D A_E 
= \frac{c}{4} \epsilon^{ABCDE} A_{A}F_{BC} F_{DE}   
\eea
We can then define the non-compactified $D=5$  theory by the action
$S=S_0 + S_{CS}$ where:
\bea
\label{SCS}
S_0 & = & \int d^5x\; L_0\; , \qquad 
S_{CS}  =  \int d^5x\; L_{CS} \;  .
\eea
The variation of the action with respect to $A_A$
generates the equation of motion:
\beq
\widetilde{e}^2\frac{\delta S}{\delta A_A} =  \partial_B F^{BA} - J^A = 0.
\eeq
We see that a conserved ``Chern-Simons current''
appears as the source term in the theory,
\beq
J^A = \frac{3c}{4}\tilde{e}^2 \epsilon^{ABCDE} F_{BC} F_{DE},
\qquad \qquad \partial^A J_A=0.
\eeq
The non-compactified action is readily seen to be gauge invariant,
provided we forbid surface terms, \ie, we
view all fields as approaching zero sufficiently
rapidly at infinity. Under a gauge
transformation:
\beq
A_A \rightarrow A_A + \partial_A \theta
\eeq
we see that $L_0$ is strictly invariant and we also have upon integrating
by parts:
\bea
\label{test}
S_{CS} & \rightarrow & S_{CS} + \frac{c}{4} \int d^5x\; \epsilon^{ABCDE}
\partial_{A}\theta \; F_{BC} F_{DE} \nonumber \\
 & = & S_{CS} - \frac{c}{4} \int d^5x\; \epsilon^{ABCDE}
\theta \partial_{A} F_{BC} F_{DE} 
\;\; = \; S_{CS}. 
\eea 

The situation changes, however, when we compactify
the theory and must accomodate surface terms.
Let us compactify the fifth
dimension, $0 \leq x^5 \leq R$.  We thus imagine that, located
at $x^5=0$ and $x^5 = R$, are surfaces (branes) denoted
respectively as $I$ and $II$, upon which we may choose to
apply various boundary conditions. 

For example, if we
apply the condition, $F_{\mu 5}|_I = F_{\mu 5}|_{II}= 0$, then we have
a ``magnetic superconducting parallel plate capacitor'' \footnote{This
is dual
to a the conventional electric superconducting parallel plate capacitor;
a magnetic superconductor is a confining phase of the theory since it admits 
electric flux tubes that would confine electric charges}.
This corresponds to an orbifold since $F_{\mu 5} =0$
can be satisfied with the gauge choice,
$\partial_\mu A_5 =0$ and $\partial _5 A_\mu =0 $.
This, in turn, implies
that $A_5$ is an odd function on the extended interval, $0\leq x^5 \leq 2R$,
while $A_\mu$ is even, corresponding to the normal orbifold configurations
for a theory compactified on $S_1/Z_2$.
The advantage of stating boundary conditions
in the ``capacitor'' language is that they are 
then manifestly gauge
invariant, \ie, orbifolding does not
break gauge invariance but, rather, corresponds to a particular gauge choice
for a physical capacitor system.

If we now repeat the check of gauge invariance, we see
that there is a surface term generated on each of the branes
that follows from performing the integration by parts:
\bea
\label{test2}
S_{CS} & \rightarrow & S_{CS} + \frac{c}{4}\int_{II} d^4x\; 
\theta(R)\; \epsilon^{\mu\nu\rho\sigma}
F_{\mu\nu} F_{\rho\sigma}(R) 
-
\frac{c}{4}\int_{I} d^4x\;
\theta(0)\; \epsilon^{\mu\nu\rho\sigma}
F_{\mu\nu} F_{\rho\sigma}(0)\;.
\eea
The theory is no longer gauge invariant since the action has
shifted by the two surface terms that take the form
of anomalies. We refer to these terms as the ``CS anomalies.''
If, however, the branes contain additional matter fields with anomalous 
gauge currents,  then they can cancel the above
CS anomalies
on the branes, and the overall theory, bulk plus branes, 
becomes gauge invariant.  

Let us presently assume that we have arranged a generic
mechanism of cancelling the CS anomaly. 
Since we are compactifying the theory on $0\leq x^5 \leq R$, it
is useful to put the CS term into a form that consists of
two terms, one that isolates $A_5$ and another that isolates $\partial_5$.
It is readily seen that $L_{CS}$ can be written as:
\beq
\label{CS0}
L_{CS} =  \frac{3c}{4} \epsilon^{\mu\nu\rho\sigma} A_{5}F_{\mu\nu} F_{\rho\sigma}
+{c} \;\epsilon^{\mu\nu\rho\sigma} (\partial_{5}A_{\mu}) A_\nu F_{\rho\sigma}\;
.
\eeq
To write $L_{CS}$ in this form we have discarded 
total divergences in the $D=4$ theory that do not affect
the physics.
In the compactified case our problem is to bring the CS term into a manifestly
gauge invariant form. We will see that this is possible for all massive
modes, but not for the zero
mode. 

Consider a Wilson line that emanates from, \eg, brane I, $ x^5 =0$, 
toward an arbitrary point in the bulk, $x^5 = y$: 
\beq
U(y) = \exp \left( i \int_0^y dx^5 A_5(x^5)\right)
\eeq
and we have:
\beq
\partial_y U = iA_5(y)U
\eeq
Using the Wilson line as a gauge transformation, we have:
\beq
\label{gt}
A_A \rightarrow A_A + iU^\dagger \partial_A U
\eeq 
and we thus see that:
\beq
A_5 \rightarrow A_5(y) + iU^\dagger \partial_y U =
A_5(y) - \partial_y \int_0^y dx^5 A_5(x^5) = 0
\eeq
Assuming the cancellation of CS anomalies occurs with some
generic matter fields, this gauge transformation 
thus annihilates the first
term of eq.(\ref{CS0}), and takes the CS term into the form
\footnote{The ``annihilation'' of the first term 
of eq.(\ref{CS0}) can be seen
to occur in detail upon performing the gauge transformation
for any particular matter anomaly cancellation with the CS anomaly. 
It involves, however,
both the first ($A^5$), as well as the second
($\partial_5$), terms of eq.(\ref{CS0}). The second term yields
$\partial_5\partial_\mu A^5$ terms that must be integrated
by parts, and the cancellation is then just the full matter anomaly
and CS term anomaly cancellation.}:
\beq
\label{S1}
L_{CS} = 
{c} \;\epsilon^{\mu\nu\rho\sigma} (\partial_{5}B_{\mu}) 
B_\nu F_{B\rho\sigma}
\eeq
where we define the gauge transformed $A_\mu$'s as $B_\mu$'s,
\beq
\label{Stu}
B_\mu = A_\mu - \partial_\mu \int_0^y A_5 dx^5; \qquad
F_{B\mu\nu} = \partial_\mu B_\nu - \partial_\nu B_\mu .
\eeq
Thus, the resulting Chern-Simons action takes the form:
\bea
\label{S2}
S_{CS} & = &
{c}\;\epsilon^{\mu\nu\rho\sigma} \int d^4 x \int_0^R dy\; (\partial_y B_{\mu}) 
B_\nu F_{B\rho\sigma} .
\eea
Eq.(\ref{S2}) is the form we desire. As we'll see
below, the $B_\mu$ field
will lead to gauge invariant (``Stueckelberg'') combinations for each
massive KK-mode in the compactified theory when we do the mode expansion.
The massless zero mode (photon) gauge field, however, will  appear
explicitly in the result, and the full zero mode gauge invariance
requires the addition of the matter effects (\eg, the triangle diagrams).
Note the $\partial_y$ in the integrand which leads to the breaking
of KK-parity. This is associated with $\epsilon_{\mu\nu\rho\sigma}$
so that, overall, only the product
of KK-parity and space-time parity is conserved.

We now turn to QED with chiral electrons
as a means of providing the anomaly cancellation
on the branes.
An alternative theory with axions on the branes is developed in Appendix A.

\section{$D=5$ QED, Orbifold Compactification}

\subsection{Chiral Fermions}

Consider QED in $D=5$ on an orbifold
with periodic domain $0 \leq x^5 \leq R$. We place
electrons on the boundary branes located
at $x^5 =0$ and $x^5=R$. The electrons
are chiral, with $\psi_L$ ($\psi_R$) on the left-brane, $I$ at $x^5 =0$
(right-brane, $II$, at $x^5=R$). 
These fermions have anomalies on their respective branes.
These matter anomalies will cancel the CS anomalies 
provided the coefficient
$c$ takes on a  special value dictated by
the fermionic anomalies. We first establish this special value
of $c$.

\begin{figure}[t]  
\vspace{4.5cm}  
\includegraphics{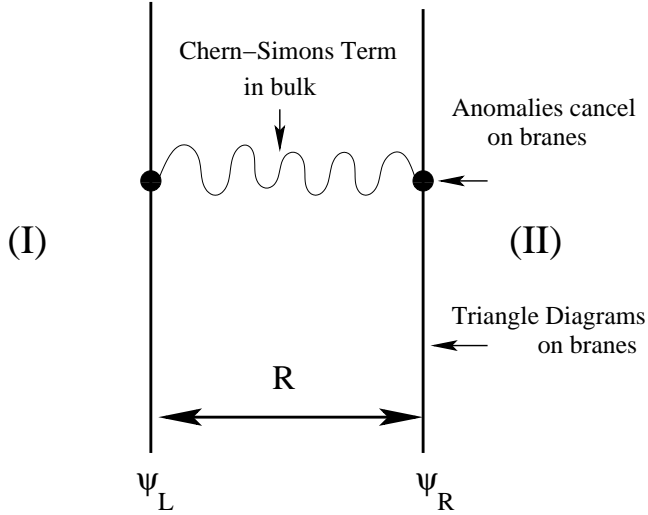}  
\vspace{3.5cm}  
\caption[]{
\addtolength{\baselineskip}{-.3\baselineskip}  
Orbifold with split, anomalous fermions (electrons).
$\psi_L$ ($\psi_R$) is attached to the $D=4$ left-brane, $I$ (right-brane,
$II$).
Gauge fields propagate in the $D=5$ bulk, which has a compactification
scale $R$. The bulk contains a Chern-Simons term, and the branes
produce triangle diagram amplitudes in the effective action. The anomalies
from the Chern-Simons term cancel the anomalies from the triangle diagrams
on the respective branes so the overall theory is anomaly free. }  
\label{dirac2}  
\end{figure}

This model has the same divergence structure
for the fermion loops as does ordinary QED in $D=4$.  
The only nontrivial new bulk
interaction is the topological CS term. It is not
hard to verify that the topological CS term
is not subject to renormalization at the one  
loop level, from diagrams involving internal
gauge fields in the continuum $D=5$ theory. 
This non-renormalization happens because
the CS-term is a topological invariant.   
This probably holds to all orders in the $D=5$ perturbation
theory, and is related
to the $D=4$ Adler-Bardeen non-renormalization theorem for the anomaly.
The CS term does renormalize the
bulk kinetic term with a quadratic divergence, 
and it may induce additional non-topological
counterterms with factors of $\tilde{e}^2$. This is the usual
problem of a $D=5$ theory, so we imagine some kind of
UV cut-off, such as an overarching string theory as the
UV completion. 

The full
action of the theory is defined as,
\beq
S=S_0 + S_{CS} + S_{branes}
\eeq
where the bulk kinetic term action for the theory is:
\bea
S_0 & =& -\frac{1}{4\widetilde{e}^2}\int_0^R dy\int d^4 x F_{\mu\nu}F^{\mu\nu}
-\frac{1}{2\widetilde{e}^2}\int_0^R dy\int d^4 x
F_{\mu 5}F^{\mu 5}.
\eea
The fermionic matter action on the branes is:
\bea
S_{branes} & = &   \int_I d^4x \;\bar{\psi}_Li\slash{D}_L\psi_L 
+  \int_{II} d^4x \;\bar{\psi}_Ri\slash{D}_R\psi_R
\eea
where: 
\beq
D_{L\mu} = \partial_\mu - iA_\mu(x_\mu, 0)\; ,
\qquad \qquad D_{R\mu} = \partial_\mu - iA_\mu(x_\mu, R)
\eeq
and
\beq
\psi_L = \frac{1-\gamma^5}{2} \psi \; , \qquad
\qquad
\psi_R = \frac{1+\gamma^5}{2} \psi
\eeq
We have thus ``split'' QED into two chiral 
theories living on distinct branes, $I$ and $II$. 
The $\psi_L$ and $\psi_R$
chiral projections are key ingredients of
the theory.
This structure can, of course, come
about if there is a thin domain wall (kink) at $x^5=0$ and an
anti-domain wall (anti-kink) at $x^5=R$, where $\psi_L$ and $\psi_R$
are then the fermionic zero modes.
We presently have no bulk propagation of the fermions, 
and this can be engineered
if the fermions have a very large Dirac mass in the bulk
away from the domain walls.  

The chiral electrons 
have anomalies on their respective branes. 
Consider a gauge transformation in the bulk:
\beq
\label{gt00}
A_A(x_\mu, y) \rightarrow A_A(x_\mu, y)
+ \partial_A \theta (x_\mu, y)
\eeq
and we therefore have:
\bea
\label{shift0}
S_{branes} & \rightarrow  & 
S_{branes}  + \int_I d^4x \;\bar{\psi}_L\gamma_\mu\partial^\mu\theta\psi_L (x_\mu, 0)
+  \int_{II} d^4x \;\bar{\psi}_R\gamma_\mu\partial^\mu \psi_R(x_\mu, R)
\nonumber \\
& \rightarrow  &
S_{branes}  -
\int_I d^4x \;\theta (x_\mu, 0) \partial_\mu J_L^\mu
 - \int_{II} d^4x \; \theta (x_\mu, R) \partial_\mu J_R^\mu
\eea
where $J_{L,R}^\mu =\bar{\psi}\gamma_\mu \psi_{L,R}$.
Note that this 
can be induced by a gauge transformation
of the electrons on the branes, 
\beq
\psi_L \rightarrow \exp(i\theta (x_\mu, 0))\psi_L \; , \qquad 
\psi_R \rightarrow \exp(i\theta (x_\mu, R))\psi_R \;.
\eeq 
At the quantum loop
level this transformation is anomalous, the currents
are not conserved. We can view
it as generating the
Noether terms on the branes of eq.(\ref{shift0}) through
the fermionic functional measure of the path integral
\cite{fujikawa}.
To proceed, we must determine the forms of the anomalies on
the branes, and this requires care.

\subsection{Consistent Anomalies on the Branes}

Consider the theory of a single Weyl spinor in $D=4$:
\beq
\label{tt0}
S = \int d^4x \; \bar{\psi}_L(i\slash{\partial}+\slash{A}_L)\psi_L
\eeq
The theory is anomalous and the anomaly is
unambiguously determined as:
\bea
\label{bardeen100}
\partial_\mu J_L^\mu & = & -\frac{1}{48\pi^2}
F_{A_L}^{\mu\nu}\widetilde{F}_{{A_L}\mu\nu}\; , \qquad 
J^\mu_{L}=\bar{\psi}\gamma^\mu\psi_L\; ,
\qquad
\makebox{and} \qquad
\widetilde{F}_{\mu\nu} = \half \epsilon_{\mu\nu\rho\sigma}{F}^{\rho\sigma}. 
\eea
Moreover, if we have a pair of Weyl spinors in $D=4$:
\beq
\label{t0}
S = \int d^4x \;\left[ \bar{\psi}_L(i\slash{\partial}+\slash{A}_L)\psi_L + 
\bar{\psi}_R(i\slash{\partial}+\slash{A}_R)\psi_R \right]
\eeq
then we can treat these fields symmetrically, each as in eq.(\ref{tt0}).
The anomalies are given implicitly in Bardeen's  paper
for the massive case, \cite{bardeen}, and 
we confirm these results in Appendix C for the massless Weyl spinor
theory of eq.(\ref{tt0}:
\bea
\label{bardeen10}
\partial_\mu J_L^\mu & = & -\frac{1}{48\pi^2}
F_{A_L}^{\mu\nu}\widetilde{F}_{{A_L}\mu\nu}; \qquad 
\partial_\mu J_R^\mu  =  \frac{1}{48\pi^2}
F_{A_R}^{\mu\nu}\widetilde{F}_{A_R\mu\nu} \;\;\;\;\;
\eea
where: 
\beq
J^\mu_{L}=\bar{\psi}\gamma^\mu\psi_L,
\qquad \qquad
J^{\mu}_{R}=\bar{\psi}\gamma^\mu\psi_R\; .
\eeq
If we construct the vector and axial vector currents,
$J=J_L+J_R$, and $J^5=J_R-J_L$, and we define
define $A_R=V+A$ and $A_L=V-A$, we have by simple algebra:
\bea
\label{bardeen11}
\partial_\mu J^\mu & = & \frac{1}{12\pi^2}
F_V^{\mu\nu}\widetilde{F}_{A\mu\nu} 
\qquad 
\partial_\mu J^{5\mu}  =  \frac{1}{24\pi^2}\left(
F_V^{\mu\nu}\widetilde{F}_{V\mu\nu} + F_A^{\mu\nu}\widetilde{F}_{A\mu\nu}
\right) 
\eea
This latter result is the form quoted in 
eq.(44) of  \cite{bardeen}. One should pay particular
attention to the coefficients in eq.(\ref{bardeen100}),
eq.(\ref{bardeen10}) and eq.(\ref{bardeen11}).  

This
form of the vector and axial-vector current anomalies 
is known as the {\em ``consistent anomaly,''} as it is consistent with
the direct calculation of the triangle diagrams of the Weyl spinors.
As stated above, the form of the anomaly for a 
theory containing {\em only} a single
pure left-handed Weyl fermion  
is unambiguous, given by the first expression
in eq.(\ref{bardeen10}). There is no ambiguity when there
is only one gauge field $A_L$ and there is no nonvanishing
counterterm (a term of the 
form $\epsilon_{\mu\nu\rho\sigma}A_L^\mu A_L^\nu \partial^\rho A_L^\sigma$
is zero)
that can be added to the theory to redefine the anomaly. 

The form in eq.(\ref{bardeen11})
is just the sum of  left-handed and right-handed
Weyl fermion consistent anomalies, and is referred to as the ``left-right
symmetric
anomaly.'' However, taken together with both the 
left-handed and right-handed Weyl spinors
and two gauge fields, $A_L$ and $A_R$, there is now an ambiguity
in the form of the anomaly. We now have the freedom
to introduce counterterms such as  
$\epsilon_{\mu\nu\rho\sigma}A_L^\mu A_R^\nu \partial^\rho A_L^\sigma$, \etc,
and these can modify the form of the anomaly.
We can now force {\em  the vector current to be conserved by
adding to the lagrangian a particular counterterm.}

To see this, consider a term in the action of the form:
\beq
\label{WZ}
S' =  \frac{1}{6\pi^2}
\int d^4x \; \epsilon_{\mu\nu\rho\sigma}A^\mu V^\nu \partial^\rho V^\sigma
\eeq
This term is unique, having even parity and nonvanishing.
Upon variation wrt $V$ or $A$ it
adds corrections to the vector and axial currents:
\bea
\label{bardeen12}
\frac{\delta S' }{\delta V_\mu} & = & \delta J^\mu  = 
-\frac{1}{3\pi^2}\epsilon_{\mu\nu\rho\sigma}A^\nu \partial^\rho V^\sigma
+\frac{1}{6\pi^2}\epsilon_{\mu\nu\rho\sigma}V^\nu \partial^\rho A^\sigma
\nonumber \\
\frac{\delta S' }{\delta A_\mu} & = & \delta J^{5\mu}  = 
\frac{1}{6\pi^2}\epsilon_{\mu\nu\rho\sigma}V^\nu \partial^\rho V^\sigma
\eea
The full currents, $\tilde{J} = J + \delta J$,
now satisfy:
\bea
\label{bardeen13}
\partial_\mu \tilde{J}^\mu & = & 0 \;,
\qquad \qquad \partial_\mu \tilde{J}^{5\mu}  = \frac{1}{8\pi^2}\left(
F_V^{\mu\nu}\widetilde{F}_{V\mu\nu} + \frac{1}{3}
F_A^{\mu\nu}\widetilde{F}_{A\mu\nu}
\right). 
\eea
This is called the ``{\em covariant}'' form of the anomaly.
The theory is now invariant, and operators transform
covariantly with respect
to the {\em vector} gauge symmetry.
We thus see that the coefficient of the $F_V\tilde{F}_V$ in the 
divergence of the axial current now corresponds to 
Adler's result in QED \cite{adler}. The vector current
is conserved even though there is an axial vector
background field. 
Thus, Adler's coefficient of the anomaly \cite{adler} arises
as a mixture of the Weyl fermion (consistent) anomaly and the
counterterm. In fact, for the $F_V\tilde{F}_V$ part,
the Adler coefficient is $1/3$ $\times$ (triangle diagrams) plus 
$2/3$ $\times$ (counterterm). These coefficients are often
confused in the literature when authors incorrectly assume
that the left-handed Weyl current 
has an anomaly coefficient that, in magnitude, is half that of Adler's result.

One might wonder what kind of UV completion theory
leads to this counterterm in the effective action. 
In fact, as we see below (and in Appendix B), this counterterm is
just the Chern-Simons term expressed as  $D=4$ effective interaction,
arising from our $D=5$ theory,
when we truncate on the zero-mode and first KK-mode.

In our present situation, therefore,
a theory with the spatial delocalization 
of the chiral fermions where anomalous
representations are placed  on  distinct branes in $D=5$,
dictates the use of the {\em consistent anomaly} on each brane.
We have the correspondence
$B_\mu(0) = A_L = V_\mu-A_\mu $ and $B_\mu(R) = A_R = V_\mu+A_\mu $, hence
the anomalies take the form on the branes: 
\bea
\label{bardeen4}
\partial_\mu J_L^\mu & = & -\frac{1}{48\pi^2}
F^{\mu\nu}(0)\widetilde{F}_{\mu\nu}(0) \qquad\makebox{(brane I)}\nonumber \\
\partial_\mu J_R^\mu & = & \frac{1}{48\pi^2}
F^{\mu\nu}(R)\widetilde{F}_{\mu\nu}(R) \;\;\;\;\; \makebox{(brane II)}
\eea
From eq.(\ref{shift0}), 
eq.(\ref{bardeen4})  the shift in the action
under the gauge transformation,
\bea
\label{fermions}
\psi_L(x_\mu) & \rightarrow & \exp(i\theta(0,x_\mu))\psi_L(x_\mu)  \nonumber \\
\psi_R(x_\mu) & \rightarrow & \exp(i\theta(R,x_\mu))\psi_R(x_\mu) 
\eea
thus takes the form:
\bea
S_{branes} & \rightarrow  &
S_{branes}  
+
\;\frac{1}{48\pi^2}\int_I d^4x \;\theta (x_\mu, 0)
F^{\mu\nu}\widetilde{F}_{\mu\nu}(0) \
-\;\frac{1}{48\pi^2}\int_{II} d^4x \; \theta (x_\mu, R)
F^{\mu\nu}\widetilde{F}_{\mu\nu}(R)
\nonumber \\
\eea
On the other hand,
the CS term produced, under
the gauge transformation, the result of
eq.(\ref{test2}) on the boundaries: 
\bea
S_{CS} & \rightarrow & S_{CS} -\frac{c}{2} \int_I d^4 x \; \theta (x_\mu, 0)
F^{\mu\nu}\widetilde{F}_{\mu\nu} 
+\frac{c}{2}\int_{II} d^4 x \; \theta (x_\mu, R)
F^{\mu\nu}\widetilde{F}_{\mu\nu}
\eea
(note the factor of $1/2$ from the introduction of $\widetilde{F}$).
The cancellation of the
anomalies therefore requires: 
\beq
\label{coeff}
c = \frac{1}{24\pi^2}
\eeq
Our discussion may have been somewhat tedious, but
this result is essential to the correct implementation of
the CS-term in a higher dimensional theory.
This is one of several ways 
to obtain the quantization of the coefficent of
the CS-term in $D=5$. With multiple copies
of boundary fermions we can have integer multiples of $c$. 
We caution the reader
that different configurations of boundary branes, or
chiral fermions in the bulk, can lead to differing results 
for $c$ in different domains of the extra dimension
(\eg, see the discussion
in the non-abelian  case below eq.(\ref{nonabcase})).
For the configuration of a physical domain $[0,R]$ with a pair of
boundary branes $I$ and $II$ the result of  eq.(\ref{coeff})
is the correct coefficient of the Chern-Simons term in the $U(1)$
and non-abelian cases.

\subsection{Mode Expansion}

Let us now consider the compactification in flat space-time.
For the orbifold (magnetic superconducting branes) 
$A_\mu$ is defined as an even function
in the doubled interval $[0,2R]$
and $A^5$ is odd. The physical extra dimension
spans the interval $[0,R]$, which
dictates the normalization
of the fields. We perform a conventional mode expansion for
the KK-mode tower of gauge fields:
\bea
\label{mode}
A_\mu^0 (x,y) & = & \sqrt{\frac{1}{ R}}\widetilde{e} A_\mu^0 (x) \nonumber \\
A_\mu (x,y) & = & \sum_{n=1}^\infty (-1)^n\sqrt{\frac{2}{ R}} \widetilde{e} \cos(n\pi y/R)A_\mu^n (x)  \nonumber \\
A_5(x,y) & = & \sum_{n=1}^\infty (-1)^{n+1}\sqrt{\frac{2}{ R}} \widetilde{e} 
\sin(n\pi y/R) A_5^n (x)
\eea
The sign conventions, $(-1)^n$, are designed so that the
$A_\mu^n$ ($B_\mu^n$; see below) with $n$ odd 
couple with a positive sign to the axial current,
$\bar{\psi}\gamma^5\psi$.

The kinetic terms contained in $S_0 \equiv S_1+S_2$ become:
\bea
S_1 & = & -\frac{1}{4\widetilde{e}^2}\int_0^R dy\int d^4 x\; F_{\mu\nu}F^{\mu\nu}
 =  -\frac{1}{4}\sum_{n}\int d^4 x\; F^n_{\mu\nu}F^{n\mu\nu}
 \eea
 \bea
\label{massterm}
S_2 & = & \frac{1}{2\widetilde{e}^2 }\int_0^R dy\int d^4 x\; F_{\mu 5}F^{\mu 5}
 =  \half \sum_{n=1}M_n^2\int d^4 x \; B_\mu^n B^{n\mu}
\eea
where the $B^n_\mu(x)$ ( $A^n_\mu(x)$) without
the argument $y$ are $D=4$ fields:
\bea
\label{gi}
M_n & = & {n\pi}/{R}\; ; 
\qquad
B_\mu^n  =  A_\mu^n + \frac{1}{M_n}\partial_\mu A_5^n\; ;
\qquad
 F^n_{\mu\nu}  \equiv  \partial_\mu B^n_\nu - \partial_\nu B^n_\mu.
\eea
We thus observe  that the gauge field mass term of eq.(\ref{massterm})
is  manifestly gauge invariant. That is, it is automatically  
expressed in terms of the  $B_\mu^n$, which are 
``Stueckelberg fields.'' The Stueckelberg fields
are combinations of transverse and longitudinal gauge fields
that are manifestly gauge invariant, \ie, if we
shift $A^n_\mu \rightarrow A^n_\mu + \partial_\mu \theta^n$
then we can also shift $A_5^n \rightarrow A_5^n -M^n\theta^n$ and
we see that $B_\mu^n$ is invariant. The $B_\mu^n$ fields
have the same  mode expansion as the $A_\mu^n$ in eq.(\ref{mode}).

 The physical value of the electric charge
follows by considering the
zero mode component of a coupling $A_\mu J^\mu$, 
where $J^\mu$ is the vector current on the branes
(sum of left current on $L$ and right current on $R$),
and we see that:
\beq
e = \widetilde{e}/\sqrt{R} \equiv e_0
\eeq
Note that $e$ is dimensionless, since $\widetilde{e}$
has dimensions of $M^{-1/2}$.  Likewise,
if we consider a transverse KK-mode coupling to a 
current on the brane, $A_\mu^n J^\mu $
we see that the coupling 
differs by a normalization factor of $\sqrt{2}$, and we
thus define:
\beq
e' = \sqrt{2}\widetilde{e}/\sqrt{R} = \sqrt{2} e \equiv e_n\;\;\;(n\neq 0)
\eeq
The $B^n_\mu$ couple as:
\bea
& & \bar{\psi}\gamma_\mu \psi_L\sum_n (-1)^n e_n B^{n\mu} 
 + \bar{\psi}\gamma_\mu \psi_R\sum_n e_n B^{n\mu}
 \nonumber \\
 & = & \bar{\psi}\gamma_\mu \psi\left(\half \sum_n (1+(-1)^n)B^{n\mu} \right) 
 + \bar{\psi}\gamma_\mu\gamma^5 \psi\left(\half\sum_n  (1-(-1)^n)B^{n\mu} \right)
\eea

We now turn to the Chern-Simons term, eq.(\ref{S2}) with
the quantized coefficient of eq.(\ref{coeff}). 
We substitute the mode expansion: 
\bea
S_{CS}   & = & \frac{1}{24\pi^2} \int_0^R dy\int d^4 x 
\;\epsilon^{\mu\nu\rho\sigma}
(\partial_{y}B_{\mu}) B_\nu F_{\rho\sigma}
\nonumber \\
   & \equiv & \frac{1}{12\pi^2}
\sum_{nmk}\int d^4 x \;
(e_ne_me_k) c_{nmk}(B^n_{\mu} B^m_\nu \widetilde{F}^{k\mu\nu})
\eea
The structure constants, $c_{nmk}$, are determined by performing
the wave-function
overlap integrals in the bulk, which is straightforward:
\bea
\label{ccc}
c_{nmk} & = & (-1)^{(k+n+m)} \int_0^1 dz\; 
\partial_z [\cos(n\pi z) ]\cos(m\pi z)\cos(k\pi z)\nonumber \\
& = & 
\frac{n^2(k^2+m^2-n^2)\left[(-1)^{(k+n+m)}-1 \right]}{
(n+m+k)(n+m-k)(n-k-m)(n-m+k)} 
\eea
Note that in particular cases of interest to us
these reduce to:
\bea
\label{ccc2}
c_{nm0} & = & c_{n0m} = 
-\frac{n^2}{n^2-m^2}\left[(-1)^{n+m}-1\right] 
\nonumber \\
c_{0nm} & = & c_{000} = 0 \nonumber \\
c_{n00} & = & 
\left[1-(-1)^{n}\right].
\eea
The selection rules for KK-mode production and decay 
can almost be inferred from these
results, but the effects of the matter fields must also
be incorporated.  For example, while the CS-term appears to allow
a KK-mode decay to two zero modes, since $c_{n00}$ is nonzero, we actually
find in Section IV 
that this is completely cancelled by the triangle diagrams in
the $m^2 > M_n^2$ limit, (while it remains
allowed in the case of axions on the branes).  The decay of an odd (even)
KK-parity mode to an even (odd) KK parity mode plus a zero mode,
through the nonzero $c_{nm0}$ is, however, allowed when the 
triangle diagram effects are included.  

The effective action for the full theory 
can now be written. 
It is again convenient to reabsorb the coupling constants into the
gauge fields, and write the effective action 
in the following compact form as an effective $D=4$ theory:
\bea
\label{effective2}
S_{full} & = & \int d^4 x\;\bigl [ \;\bar{\psi}
(i\slash{\partial}+ \slash{V} +
\slash{{\cal{A}}}\gamma^5-m)\psi 
+ \frac{1}{12\pi^2} \sum_{nmk}c_{nmk}B^n_\mu B^m_\nu 
\widetilde{F}^{k\mu\nu}
\nonumber \\ & & 
-\frac{1}{4e^2} F^0_{\mu\nu}F^{0\mu\nu}
-\frac{1}{4e'{}^2}\sum_{n\geq 1} F^n_{\mu\nu}F^{n\mu\nu} 
+\sum_{n\geq 1} \frac{1}{2e_n^2}M_n^2 B_\mu^n B^{n\mu}  \bigr ]
\eea
where:
\bea
V_\mu & = & \sum_{n\; even}B_\mu^n, \;\;\;\qquad
{\cal{A}}_\mu  =  \sum_{n\; odd}{B}_\mu^n
\eea
and the photon zero mode is defined as:
$A_\mu \equiv B_\mu^0.$
Sums over even $n$ now
include $n=0$, unless otherwise indicated.
We caution that $S_{tree}$ must still be supplemented by the boundary
brane matter effects \ie, the
triangle diagram loops of Appendix C.

In eq.(\ref{effective2}) we have supplemented the action with
an  electron mass term. The electron mass has to be viewed
as arising from a bilocal bulk term in the parent theory of the form:
\beq
m\bar{\psi}_L(x_\mu,0) W \psi_R(x_\mu,R) + h.c., \qquad \qquad 
W = \exp(i\int_0^R A_5(x_\mu,x_5) dx^5)
\eeq 
This is gauge invariant in the full $D=5$ theory owing
to the Wilson line, $W$.  When we
perform the gauge transformation of 
eq.(\ref{gt}) the Wilson line becomes $W= U(0)W U(R)^\dagger = 1  $
and the mass
term goes into the Dirac form, $-m\bar{\psi}\psi$,
as displayed in eq.(\ref{effective2}).
The full set of gauge transformations on the branes thus
form a $U(1)_L\times U(1)_R$ gauged chiral lagrangian.
Since the $B_\mu^n$ axial fields are in the Stueckelberg form,
there is no conflict with the gauge invariance of the 
electron mass term. Moreover, the vectorial gauge transformations, \ie, those
with $\theta(0) = \theta(R)$
commute with the electron mass term.

The KK-modes with even (odd) $n$ must now 
 be interpreted as vectors, with $J^P=1^-$ (axial-vectors,
with $J^P=1^+$).  The $\epsilon$ symbol
has locked the internal KK-mode parity of the $x^5$ wave-functions
in the bulk to the parity of space-time.  Put another way,
the CS-term is explicitly violating the notion of independent
symmetries of KK-mode parity and space-time parity. 
All of the massive $B^n_\mu$ are seen to be gauge invariant 
in the sense of  Stueckelberg fields,
\ie, they appear in the gauge invariant combinations as written in
eq.(\ref{gi}). 

Note that if we truncate the theory on the zero mode $B^0$ and first KK-mode,
$B^1$, the CS term goes into the form:
\beq
\frac{1}{12\pi^2} c_{100}B^1_\mu B^0_\nu 
\widetilde{F}^{0\mu\nu} \; = \; \frac{1}{6\pi^2}\epsilon^{\mu\nu\rho\sigma} 
A_\mu V_\nu  {\partial}_{\rho}V_{\sigma}
\eeq
where the $c_{111}$ term vanishes by antisymmetry.
This is precisely the same form as the counterterm of eq.(\ref{WZ}),
and correspondingly the full vector and axial vector currents have
covariant anomalies as in eq.(\ref{bardeen13}) 
Using a remarkable
identity amongst the $c_{nmk}$ structure constants, obtained
in eq.(\ref{dd}), 
the full current and anomaly structure for the tower
of KK-modes is derived in Appendix B. This yields the {covariant} form 
of the anomaly for the tower of KK modes, and the vector zero-mode
current is indeed conserved.

\subsection{The Full Action Including Matter Effects}

We now supplement the full tree action of eq.(\ref{effective2})
with the contributions of the matter fields. We do
this by integrating out the matter fields.
The gauge field part of the full action 
now takes the form:
\bea
\label{effective3}
S_{tree} & = & \int d^4 x\;\bigl [ 
\frac{1}{12\pi^2} \sum_{nmk}\bar{c}_{nmk}B^n_\mu B^m_\nu 
\widetilde{F}^{k\mu\nu}
\nonumber \\ & & 
-\frac{1}{4e^2} F^0_{\mu\nu}F^{0\mu\nu}
-\frac{1}{4e'{}^2}\sum_{n\geq 1} F^n_{\mu\nu}F^{n\mu\nu} 
+\sum_{n=0} \frac{1}{2e_n^2}M_n^2 B_\mu^n B^{n\mu}  \bigr ]
\eea
where the ${c}_{nmk}$ have been replaced by the $\bar{c}_{nmk}$.

In Appendix A we show
that with axions as matter fields the CS term is unmodified:
\beq
\bar{c}_{nmk} = {c}_{nmk} \qquad \qquad \makebox{(axions)}.
\eeq
In this case the photon has acquired a mass, and
we include a nonzero $M^2_0$ term. There is also a
residual physical axion, $\phi^{(-)}$ and terms
containing it must be included into eq.(\ref{effective3})
from  eq.(\ref{axionaction4}).  The full 
mass matrix of KK-modes further
involves small off-diagonal mixings of the massive
photon with the $n$ even KK-modes, as displayed
in eq.(\ref{axionaction4}).  These off-diagonal mixings are
negligible, but the diagonal photon mass in a unitary gauge
makes the photon itself a Stueckelberg field and 
gauge invariance is protected by shifts in the 
$\phi^{(+)} = (\phi_L-\phi_R)/2 $ combination
of the brane axions. 

The more interesting case of chiral fermions is studied in Appendix C.
The simplest case is that of the decoupled fermion, in which
$m$ is large compared to $M^a$, $M^b$ and $M^c$, and $a\rightarrow b+c$
is the exclusive tree body decay mode mediated by the CS term and anomaly.
We obtain for the boundary term
an effective operator, ${\cal{O}}_3$, describing
the 3-gauge boson amplitude of the triangle diagrams from Appendix C,
in eq.(\ref{an10}) (with coupling constants restored):
\beq
\label{calo3}
{\cal{O}}_3 =  -\frac{1}{12\pi^2}\epsilon^{\mu\nu\rho\sigma} 
\sum_{nmk} (e_n e_m e_k)a_{nmk}B^n_\mu B^m_\nu\partial_\rho B_\sigma^k 
\eeq
where:
\beq
\label{acoeff}
a_{nmk} = \half (1 - (-1)^{n+m+k})(-1)^{m+k}
\eeq
Note that this has the form of the counterterm that mediates consistent
and covariant anomalies when truncated on the lowest modes.
In this case, we therefore have:
\beq
\bar{c}_{nmk} = {c}_{nmk} - a_{nmk} \qquad \makebox{(massive spinors)}
\eeq
hence we can write:
\beq
\label{barc}
\bar{c}_{nmk} = \left[(-1)^{(k+n+m)} -1\right]
\left(\frac{n^2(k^2+m^2-n^2)}{
(n+m+k)(n+m-k)(n-k-m)(n-m+k)} + \half (-1)^{m+k}\right)
\eeq
By adding these terms
we are decoupling the fermions 
and the effective action becomes purely bosonic.


\section{New Physics from the Chern-Simons Term}

From the CS term in eq.(\ref{effective3}) we deduce
the Feynman rule for a vertex as shown in Fig. 2
for the process $B^a \rightarrow B^b + B^c$:
\bea
\label{effective5}
T_{CS} & = & -\frac{ee'{}^2}{12\pi^2} \big[
(-\bar{c}_{abc}+\bar{c}_{bac}+\bar{c}_{bca}-\bar{c}_{cba})[B]
+ (\bar{c}_{acb}-\bar{c}_{cab}+ \bar{c}_{bca}-\bar{c}_{cba})[A]
\big]
\eea
where $[A]$ and $[B]$ are (as in Appendix C):
\beq
[A] = \epsilon^{\mu\nu\rho\sigma}
\epsilon^a_\mu \epsilon^b_\nu\epsilon^\gamma_\rho k^\sigma
\qquad
[B] = \epsilon^{\mu\nu\rho\sigma}
\epsilon^a_\mu \epsilon^b_\nu\epsilon^\gamma_\rho q^\sigma
\eeq
Here we have used momentum conservation
$ p = k + q$.
 and 
we have rescaled coupling constants back into the
interaction of eq.(\ref{effective3}).
As written, $q$ and $k$ are outgoing momenta, and
give a factor of $+i$; we also have $+i$ from $e^{iS}$
and $-1$ from the CS-term coefficient.

\subsection{Decay of KK-mode to KK-mode plus $\gamma$}

\begin{figure}[t]  
\vspace{4cm}  
\includegraphics{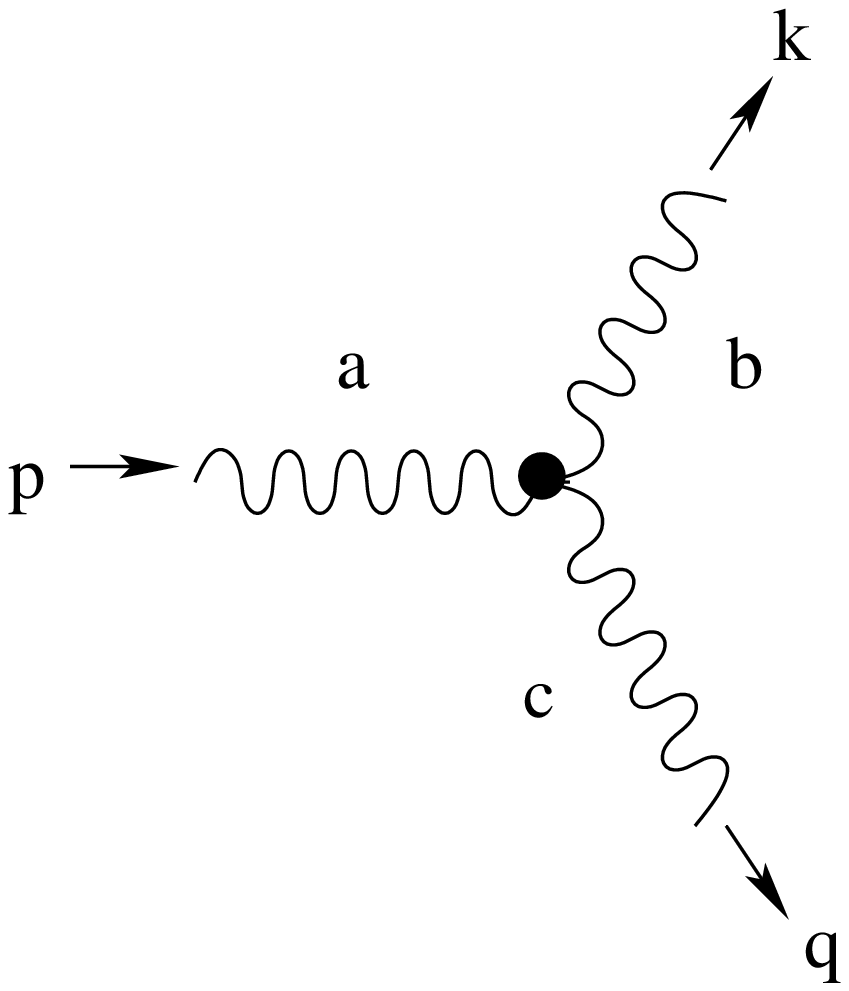}  
\vspace{1.5cm}  
\caption[]{
\addtolength{\baselineskip}{-.3\baselineskip}  
Chern-Simons term plus triangle diagrams yield a three body vertex
describing the decay of KK-modes $a\rightarrow b + c$. In the text we study
the special case where $c=0$ and $B^c$ is the photon $\gamma$.}  
\label{dirac2}  
\end{figure}  

As an example application of
the formalism, let us now
compute the tree approximation decay width of the $a$th KK-mode
into the $b$th KK-mode plus a massless zero-mode (photon).

For the $a$th KK-mode of 4-momentum $p_\mu$, polarization $\epsilon_\mu^a$,
 decaying to 
the $b$th mode of momentum $k_\mu$, polarization $\epsilon_\mu^b$
and the $\gamma$
of momentum $q_\mu$, polarization $\epsilon_\mu^\gamma$, eq.(\ref{effective3})
leads to the Feynman rule for the 3-body vertex of Fig.(2):
\bea
\label{effective5}
T_{CS} & = & -\frac{ee'{}^2}{12\pi^2} \big[
(-\bar{c}_{ab0}+\bar{c}_{ba0}+\bar{c}_{b0a}-\bar{c}_{0ba})[B]
+ (\bar{c}_{a0b}-\bar{c}_{0ab}+ \bar{c}_{b0a}-\bar{c}_{0ba})[A]
\big]
\eea
The pure CS term using the coefficients $c_{nmk}$ yields
\bea
T_{CS} & = & \frac{ee'{}^2}{6\pi^2}
 \left[
\left(\frac{a^2+2b^2}{a^2-b^2}\right)[B]-[A]
\right]
\eea
Here we have used the pure $c_{nmk}$ structure constants 
of eqs.(\ref{ccc},\ref{ccc2}) 
and we have assumed that $a+b$ is odd, as required for a nonzero result.
Note that the KK-mode masses are $M_a=\pi a/R$, so the vertex
rule can  be written as:
\bea
\label{effective5b}
T_{CS} & = & \frac{ee'{}^2}{6\pi^2}
 \left[
\left(1 + \frac{3M_b^2}{M_a^2-M_b^2}\right)[B] -[A]
\right]
\eea
As we have stressed, it is the combination of
the matter fields and the CS-term that is gauge invariant.

For the massive fields, the condition on the outgoing
polarization $p^\mu\epsilon_\mu^a$
is fixed (essentially by our gauge choice, $A^5=0$). 
This arises from the free field equation of motion of
the Stueckelberg fields, $\partial_\mu F^{a\mu\nu}
+M_a^2 B_\nu = 0$ whence $\partial_\mu B^\mu = 0$. 
However,
for the massless zero mode (photon) there is no such restriction on
the polarization.
Correspondingly under the zero mode
gauge transformation:
\beq
\label{shift}
\epsilon^\gamma_\mu \rightarrow \epsilon^\gamma_\mu + \kappa\; q_\mu
\eeq 
we see that the term $[B]$ is gauge invariant, while
$[A]$ is {\em not} invariant, undergoing a shift with
eq.(\ref{shift}). 

We now include the fermion triangle loops. 
The triangle diagrams in the large $m$ limit 
produce a simple expression for the amplitude
that has been computed in eq.({\ref{calo3}), or eq.({\ref{an10})
and thus contribute the $a_{nmk}$ coefficients
of eq.(\ref{acoeff})
to the $\bar{c}_{nmk} = {c}_{nmk}-a_{nmk}$ combination. 
For the
process of interest, $B^a\rightarrow B^b + \gamma$ we see that
the eq.({\ref{calo3}) contribution takes the form: 
\bea
\label{effective5c}
 T_L+T_R & = & \frac{e e'{}^2}{6\pi^2}\left[[A] 
 - \half(3(-1)^b -1) [B]\right] +{\cal{O}}(1/m^2)
\eea
We can check this result directly from the vertices
in eqs.(\ref{final},\ref{final2}) where we take $c=0$ and $a+b$ is then
odd. Hence,
$f_{ab0} = 2$ and $g_{ab0} = -(3(-1)^b-1)$ and eq.(\ref{final})
yields the above result.

Thus, the resulting 
full Feynman vertex rule is the sum of eq.(\ref{effective5b})
and eq.(\ref{effective5c}), and takes the form:
\bea
\label{effective6}
\bar{T} \equiv T_{CS}+T_L+T_R & = & 
 \frac{e e'{}^2}{2\pi^2}
\left(\frac{M_b^2}{M_a^2-M_b^2} - \half((-1)^b-1)\right)[B]. 
\eea
We see that the $[A]$ term, which violated the electromagnetic
gauge invariance, has miraculously cancelled, as indeed it must!
We now have a result that is fully gauge invariant: 
the longitudinal component of the zero-mode,
$\epsilon^\gamma_\mu \propto q_\mu$,
decouples from the full amplitude, since
$B\rightarrow 0$ when $\epsilon^c_\mu \rightarrow q_\mu$. 

The physical transition amplitudes for abnormal
parity, $1^+$ [$a$ odd, $b$ even], and normal parity, $1^-$ [$a$ even,
$b$ odd], $B^a$ decay thus take the  compact forms:
\bea
\label{effective6}
\bar{T}^+ & = & 
 \frac{e e'{}^2}{2\pi^2}
\left(\frac{M_a^2}{M_a^2-M_b^2} \right)[B] 
 \qquad 1^-\rightarrow 1^+ + \gamma 
\nonumber \\
\bar{T}^- & = & 
 \frac{e e'{}^2}{2\pi^2}
\left(\frac{M_b^2}{M_a^2-M_b^2} \right)[B] 
\qquad 1^+\rightarrow 1^- + \gamma
\eea
Several comments on the physical structure of the decay amplitude
are in order.
We see that if the massive gauge fields are
nearby in mass, $M_a^2 \approx M_b^2$, then the amplitude
is mainly dominated by the CS term,  eq.(\ref{effective5b}).
On the other hand, in the limit $M_b^2 << M_a^2$ the 
decay amplitude is dominated by a
coherent superposition of the Chern-Simons term
and the matter anomalies (triangle diagrams)
of eq.(\ref{effective5c}). This does not mean that the partial width
is predominantly governed by the $M_a^2 \approx M_b^2$ limit, since here
the phase space is becoming small (in fact, we'll
see that it is dominated by the decays to the lightest KK-mode
longitudinal components, since those have the smallest ``decay constants,''
\ie, the final state couples as $\sim k^\mu/M_b$ giving
an $(M_a/M_b)^2$ enhancement).
The smallest allowed
difference $M_a^2 - M_b^2$ for large $a$, is given
by $b=a-1$, whence the mass $M_a^2 = a^2\pi^2/R$ 
and mass difference is
$M_a^2 - M_{a-1}^2 \approx 2a\pi^2/{R^2}$.
In the limit
of large $a $ and $b=a-1$ we then have:
\bea
\label{effective6}
\bar{T}^+ & \approx & \bar{T}^+ \approx
 \frac{ee'{}^2a}{4\pi^2}[B] 
\eea
This growth of the amplitudes with large $a\approx b$ can
be viewed as a short distance limit.
The factor $a $ is yielding the power-law running of the  coupling
constant, $e'{}^2 a \equiv e'{}^2 (M_a) $.
In the limit $M_b^2 << M_a^2 $ the decay amplitudes 
are suppressed, and by
the anomaly (triangle diagrams) CS-term interference amplitude:
\bea
\label{effective6}
\bar{T} & \rightarrow & 
 \frac{ee'{}^2}{4\pi^2}
\left((-1)^b+1\right)[B] 
\eea
The odd-even effect here,
is a consequence of the $D=5$ overlapping
wavefunctions.
It is remniscent of a chiral flip suppression as in,
\eg, $\pi^\pm$ decay to $e\nu$, but it involves the particular
matching of the CS term with the boundary anomaly in a
way that makes the result somewhat opaque. This
can presumably vary if a different theory is taken on the branes.
For example, in the case of the brane axions as in Appendix A
there is no brane triangle diagram contribution and this 
odd-even effect is  washed out.

Turning to the partial
width calculation, 
the zero-mode gauge invariance implies, as usual,
that in
summing over final state $\gamma$ polarizations we can use
the familiar:
\beq
\sum_\lambda \epsilon^\gamma_\mu(\lambda) \epsilon^\gamma_\nu(\lambda)
= -g_{\mu\nu}
\eeq
and conveniently drop the (singular) $q_\mu q_\nu/q^2$ 
terms.  

For the heavy vector meson polarizations  the 
source-free equations
of motion, eq.(\ref{eom}), imply that
the $B_\mu^n$ (Stueckelberg) fields must obey the Lorentz gauge condition,
$\partial^\mu B_\mu^n =0$
(as usual,
the outgoing and incoming fields are treated 
as freely propagating fields, so we ignore their anomalous
source terms, which are part of the interaction generating
the transition). Thus, the polarization sums for
massive fields are:
\bea
\sum_\lambda \epsilon^a_\mu(\lambda) \epsilon^a_\nu(\lambda)
& = & -\left(g_{\mu\nu}- \frac{p_\mu p_\nu}{M_a^2}\right)
\nonumber \\
\sum_\lambda \epsilon^b_\mu(\lambda) \epsilon^b_\nu(\lambda)
& = & -\left(g_{\mu\nu}- \frac{k_\mu k_\nu}{M_b^2}\right)
\eea

Note that if we had axions as matter fields there
would be no loop correction and the original vertex, $T_{CS}$,
of eq.(\ref{effective5b}) would be the full result. 
However, then the zero mode photon would
be a massive Stueckelberg field, and we would use the polarization sums:
\bea
\sum_\lambda \epsilon^\gamma_\mu(\lambda) \epsilon^\gamma_\nu(\lambda)
& = & -\left(g_{\mu\nu}- \frac{p_\mu p_\nu}{m_\gamma^2}\right)
\eea

Squaring the amplitude and summing over the  $b$ and $\gamma$
polarizations and averaging over the $a$ polarization
yields:
\bea
<\bar{T}^2> & = &  \frac{1}{3}\left(\frac{e^3}{\pi^2}\right)^2
\left(\frac{M_a^2}{M_b^2} \right)
(M_a^2 + M_b^2) \qquad 1^-\rightarrow 1^++\gamma
\nonumber \\
<\bar{T}^2> & = &  \frac{1}{3}\left(\frac{e^3}{\pi^2}\right)^2
\left(\frac{M_b^2}{M_a^2}\right)
(M_a^2 + M_b^2) \qquad 1^+\rightarrow 1^- +\gamma
\eea
where we have set $e'{}^2 = 2 e^2$.
The former case is quite singular in the $M_b<< M_a$ limit.
This arises from the decay of the transverse $1^-$ particle
into the longitudinal $1^+$. 
The $M_b^2/M_a^2$ suppression in the $1^+\rightarrow 1^- +\gamma$
is the analogue of a chiral suppression, such
as in $\pi\rightarrow e\nu$.  

Thus, putting in phase space,
the partial decay width for $a\rightarrow b + \gamma$
in the $M_a^2 >> M_b^2$ limit, is:
\bea
\Gamma_{1^-\rightarrow 1^+\gamma} & = &  
\frac{2\alpha^3}{3\pi^3}\left( \frac{M_a^3}{M_b^2} \right)\; , 
\qquad \qquad
\Gamma_{1^+\rightarrow 1^-\gamma}  =   
\frac{2\alpha^3}{3\pi^3}M_b \; .
\eea
In the limit $\Delta M = M_a - M_b<< M_a$ we have:
\bea
\Gamma_{1^\pm\rightarrow 1^\mp\gamma} & = &  
\frac{2\alpha^3}{3\pi^3}\Delta M \; .
\eea
The most conspicuous effect is the $M_a^2/M_b^2$ enhancement.
This could be  absorbed into $\alpha'{}^2$ factors ($\alpha' = 2\alpha$) and
viewed as a power-law running of the coupling constants:
\bea
\Gamma_{1^-\rightarrow 1^+\gamma} & = &  
\frac{\alpha(0)\alpha'{}^2(M_a)}{6\pi^3}\left( \frac{M_a}{R^2 M_b^2} \right) 
\nonumber \\
\Gamma_{1^+\rightarrow 1^-\gamma} & = &  
\frac{\alpha(0)\alpha'{}^2(M_a)}{6\pi^3}\left( \frac{M_b}{R^2 M_a^2} \right)
\eea
The point of writing this latter result 
is that there is nothing particularly pathological
about the enhanced decay rates of superheavy KK modes
through the CS term, compared to the usual
pathologies of extra dimensions.  The expected order of magnitude
for such decays would have been $\sim \alpha(0) [\alpha'(M_a)]^2 M_a$
and we see that typically $R^2M_b^2= \pi^2b^2 >> 1$. Thus,
these decay widths are consistent with naive expectations.

The quadratic growth of the
width $\Gamma_{1^-\rightarrow 1^+\gamma} \propto M_a^2$ 
is presumably related to the
quadratic divergence of the two point function in the continuum
$D=5$ theory where the Chern-Simons term plays the role of the
interaction as in Fig.(2). While we have not performed
the detailed analysis, this loop presumably satisfies
a $D=5$ dispersion relation (a sum-rule in
the $D=4$ effective theory), and yields the radiative
corrections to the power-law growth of
the coupling constant.

\subsection{Zero Mode $+$ Zero Mode $\rightarrow $  KK-Mode Vanishes}

We note that the CS-term contains the vertex
describing $a\rightarrow 0 + 0$:
\bea
\label{effective13}
T_{CS} & = & -\frac{ee'{}^2}{12\pi^2} \big[
(-\bar{c}_{a00}+\bar{c}_{0a0}+\bar{c}_{00a}-\bar{c}_{00a})[B]
+ (\bar{c}_{a00}-\bar{c}_{0a0}+ \bar{c}_{00a}-\bar{c}_{00a})[A]
\big]
\eea
Consider the case in which $\bar{c}_{nmk} = c_{nmk}$
and using $c_{a00} = (1-(-1)^a)$, $c_{0mk} = 0$
we have the amplitude:
\beq
T_{CS}= -(1-(-1)^a) \frac{ee'{}^2}{12\pi^2} ([A]-[B] ) 
\eeq
This is the result for the axionic case.

For fermionic
matter fields we see from eq.({\ref{acoeff}) that the
$a_{a00} = (1-(-1)^a)/2$, $a_{0a0} = a_{00a}=(-1)^a(1-(-1)^a)/2$,
thus:
\beq
T_{L}+ T_R = (1-(-1)^a)\frac{ee'{}^2}{12\pi^2} ([A] - [B]),
\eeq
and, the combined amplitude is:
\beq
\label{cancel}
T_{CS} + T_{L} + T_{R}  = 0.
\eeq
There are no couplings of a single KK-mode to two zero-modes in the 
$m^2 > M_a^2$ limit. This is a consequence of gauge invariance
of the zero mode.

In the case of axions on the branes we would have a
zero-mode $+$ zero mode $\rightarrow $ KK-mode vertex, but
then the zero-mode is {\em not} massless. The cancellation
of eq.({\ref{cancel})
may be a general result for a massless zero-mode.
We caution that we have proved it in the large $m$ case
and the effects of massless fermions, or an off-shell
zero-mode, may be non-zero. It would be of interest
to explore the standard model with split fermions, \eg,
in which the $U(1)_Y$ anomalies are delocalized,  
to examine if processes like $Z+\gamma \rightarrow B^a$ can occur 
via mixing, or the nonzero $Z$ mass.

\section{The Non-Abelian Case}

We now consider an $SU(N)$ gauge theory 
in $D=5$, with the covariant derivative:
\beq
D_A = \partial_A - iA_A \qquad A_A \equiv A_A^a\frac{\lambda^a}{2}
\eeq
The field strength and the 
kinetic term lagrangian density are:
\bea
G_{AB} & = & i[D_A,D_B]
  =  \partial_A A_B - \partial_B A_A -i[A_A, A_B];
\qquad
 L_0 =  -\frac{1}{2\widetilde{g}^2} \Tr(G_{AB}G^{AB}),
\eea
and $1/\widetilde{g^2}$ again has dimension of $M^1$.
A gauge transformation involves a local gauge rotation,
$U = \exp(i\theta^a(x^A)\lambda^a/2)$ and the gauge
field and covariant field strength transform as:
\beq
A_A\rightarrow U^\dagger iD_A U + A_A
-\partial_A\theta^a\frac{\lambda^a}{2} 
- i\theta^a A_A^b \left[\frac{\lambda^a}{2},\frac{\lambda^b}{2}  \right]
+ \; ...
\eeq

The Yang-Mills Chern-Simons term,
also known as the ``second Chern character,''
takes the form:
\bea
\label{CSterm0}
{\cal{L}}_{CS} & = & c\;\epsilon^{ABCDE}
\Tr \Bigl ( A_A \partial_B A_C \partial_D A_E   
 - \frac{3i}{2}A_A A_BA_C \partial_D A_E 
- \frac{3}{5}A_A A_B A_C A_D A_E \Bigr ) 
\eea
This can be rewritten in a convenient
form involving gauge covariant field
strengths,
\bea
\label{CS2}
{\cal{L}}_{CS} 
& = &
\frac{c}{4}\epsilon^{ABCDE}
\Tr \Bigl (A_A G_{BC}G_{DE}   
 + i A_A A_B A_C G_{DE}
- \frac{2}{5}A_A A_B A_C A_D A_E \Bigr ) ,
\eea
It should be noted that these expressions vanish
for gauge groups having no $d$ symbols. 
(See \cite{zachos} for further discussion and references).

We now include
the Chern character into the action of a $D=5$ 
theory 
as in the the QED case: 
\bea
\label{SCS2}
S_0 & = & \int d^5x\; L_0; \qquad 
S_{CS}  =  \int d^5x\; L_{CS}  .
\eea
The variation of the action with respect to $A_A$
again generates the equation of motion:
\beq
\widetilde{e}^2\frac{\delta S}{\delta A^a_A} =  
[D_B, G^{BA}]^a - J^{aA} = 0.
\eeq
and there is again a conserved Chern-Simons current
appearing as the source term in the theory,
\bea
\label{current2}
J^a_A & = &  \frac{3c}{2}\; \epsilon_{ABCDE}
\Tr(\frac{\lambda^a}{2}\{G^{BC},G^{DE}\}).
\eea
The current explicitly requires that $SU(N)$ possess a $d$-symbol, 
hence $N\geq 3$; and it is covariantly
conserved, $[D^A,J^a_A\lambda^a/2]=0$. 

The CS-anomaly arising on boundaries under a gauge
transformation is a more complicated expression. In anomaly
matching to boundary matter fields it suffices to
keep track only of the $\Tr(dAdA)$ terms.
If we  compactify the theory as
we did in the QED case with boundary branes $I$ and $II$,
we see that, under an infinitesimal gauge transformation, 
there are surface terms:
\bea
\label{test2ym}
S_{CS} & \rightarrow & S_{CS} + c\int d^4x\; 
\theta^a(R)\; \epsilon^{\mu\nu\rho\sigma}
\Tr(\frac{\lambda^a}{2}\partial_\mu A_\nu \partial_\rho A_\sigma)\; + \;...
\nonumber \\
&  & 
\qquad \qquad  -
{c}\int d^4x\;
\theta^a(0)\; \epsilon^{\mu\nu\rho\sigma}
\Tr(\frac{\lambda^a}{2}\partial_\mu A_\nu \partial_\rho A_\sigma) + \; ...
\eea
The action has
shifted by the two  CS anomalies.

We again consider the orbifold compactification 
with branes I and II now containing, 
respectively, quarks $q_L$ and $q_R$ transforming
as $N$ under the $SU(N)$ gauge group. For our
envisioned application we view $SU(N)$ to be
the flavor symmetry, so we'll simply assume the quarks
also carry an (ungauged) color index $N_c$.

The anomalies on the branes of the chiral quarks are
again required to be the {\em consistent} Yang-Mills anomalies. 
These are
given in full form in Bardeen's paper (and we can infer
the $\Tr dAdA$ terms from Appendix C). Written
in terms of
the chiral quarks on their respective boundary branes
we have:
\bea
\label{bardeen20}
\partial_\mu J^{a\mu}_L & = & -\frac{N_c}{24\pi^2}
\epsilon^{\mu\nu\rho\sigma}
\tr\left(\frac{\lambda^a}{2}\partial_\mu A_{L\nu} \partial_\rho A_{L\sigma} \right)
+ \; ... 
\qquad 
\nonumber \\
\partial_\mu J_R^{a\mu} &  =  & \frac{N_c}{24\pi^2}
\epsilon^{\mu\nu\rho\sigma}
\tr\left(\frac{\lambda^a}{2}\partial_\mu A_{R\nu} \partial_\rho A_{R\sigma} \right)
+ \; ...  
\eea
where: 
\beq
J^{a\mu}_{L}=\bar{q}\gamma^\mu\frac{\lambda^a}{2}q_L,
\qquad \qquad
J^{a\mu}_{R}=\bar{q}\gamma^\mu\frac{\lambda^a}{2}q_R, \qquad
\makebox{and} \qquad
\widetilde{G}_{\mu\nu} = \half \epsilon_{\mu\nu\rho\sigma}{G}^{\rho\sigma}. 
\eeq
Under the $SU(N)$ flavor gauge transformation the quarks undergo the
transformation:
\beq
q_L \rightarrow \exp\left( i\theta^a(0)\frac{\lambda^a}{2} \right) q_L
\qquad
q_R \rightarrow \exp\left( i\theta^a(R)\frac{\lambda^a}{2} \right) q_L
\eeq
and the consistent anomalies produce the shifts on
the branes:
\bea
\label{test3ym}
S_{quark} & \rightarrow & S_{quark} - \frac{N_c}{24\pi^2}\int_{II} d^4x\; 
\theta^a(R)\; \epsilon^{\mu\nu\rho\sigma}
\Tr(\frac{\lambda^a}{2}\partial_\mu A_\nu \partial_\rho A_\sigma)
\nonumber \\
&  & 
\qquad \qquad  +
\frac{N_c}{24\pi^2}\int_{I} d^4x\;
\theta^a(0)\; \epsilon^{\mu\nu\rho\sigma}
\Tr(\frac{\lambda^a}{2}\partial_\mu A_\nu \partial_\rho A_\sigma) + \qquad ...
\eea
Thus, we conclude:
\beq
\label{nonabcase}
c = \frac{N_c}{24\pi^2}
\eeq
This coefficient applies in the case we have constructed
of two branes  $I$ and $II$ bounding the physical interval $[0,R]$.
One can consider a different construction in which the physical
interval is extended  to $[0,2R]$, but
the branes  $I$ and $II$ remain located at $x^5=0$ and $x^5=R$,
and we use periodic boundary conditions on all fields. This physically
corresponds to a kink $+$ anti-kink soliton with fermionic zero-modes on
$S_1$. If we impose a parity
symmetry on $[0,R] \leftrightarrow [R,2R]$ then 
we would have a CS term in the domain $[0,R]$
with coefficient $c_1= -1/48\pi^2$ and an anti-CS term in the domain
$[R,2R]$ with $c_2 = 1/48\pi^2$. In fact, 
without the parity symmetry, the domains can
have CS terms with arbitrary coefficients, $c_1$ and $c_2$. The
anomaly matching simply requires $c_1-c_2 = -1/24\pi^2$.
For concreteness we use the orbifold with physical domain
$[0,R]$ and $c_1 = -1/24\pi^2$ in the application below. 

Note that we can also introduce a Wilson line mass term for
our separated quarks of the form:
\beq
\int d^4x\; m\;\bar{q}_L(0)P\exp\left(i\int_0^R A_5 \; dx^5 \right)q_R(R) + h.c.
\eeq
This will play the role of a constituent quark mass
below when we derive the constituent chiral quark model,
and $P\exp(-i\int_0^R A_5 dx^5) = \exp(2i\tilde{\pi}/f_\pi)$
plays the role of the chiral field of mesons when we truncate
on the zero-mode of the theory. In this way, the Yang-Mills theory
of flavor can be morphed into an $SU(N)_L\times SU(N)_R$ 
chiral lagrangian of mesons, and
the CS term becomes the Wess-Zumino-Witten term, as developed
using deconstruction in ref.(\cite{zachos}).

The Chern-Simons term
may again be written in a form, as in the $U(1)$ case,
that separates the $A_5$ and $\partial_5$ terms,
\bea
\label{CS11}
{\cal{L}}_{1} = 
 \frac{c}{2}\Tr ((\partial_5A_\mu)K^\mu)
+\frac{3c}{4}\epsilon^{\mu\nu\rho\sigma}\Tr(A_5 
G_{\mu\nu}G_{\rho\sigma}),
\eea
where we find \cite{zachos}:
\bea
K^{\mu} \equiv  \epsilon^{\mu\nu\rho\sigma}
\left(iA_\nu A_\rho A_\sigma + G_{\nu\rho}A_\sigma
+ A_\nu G_{\rho\sigma}\right).
 \eea
In deriving this result, some total divergences in
the $D=4$ subspace have been discarded, which play
no role in the physics or in the anomaly matching,
as in the $U(1)$ case.
 
Again we can perform the gauge transformation that 
sets $A^5=0$ using the Wilson line $U(y)$ and defining
the ``Stueckelberg fields'' $B_\mu$ as:
\beq
U(y) = P\exp (i\int_0^{y} A_5(x^5) dx^5) \qquad\qquad
B_\mu = U^\dagger [iD_\mu, U].
\eeq
The Chern-Simons action thus becomes:
\bea
\label{CS20}
{\cal{L}}_{1} & = &
 \frac{c}{2}\epsilon^{\mu\nu\rho\sigma}\int d^4x \int_0^R dy
 \Tr \left[\partial_y B_\mu\left(iB_\nu B_\rho B_\sigma + G_{\nu\rho}B_\sigma
+ B_\nu G_{\rho\sigma}\right)\right] 
\eea
and the quark mass term goes into the 
Dirac form $m\bar{q}_L q_R+h.c. = m\bar{q} q $.

As we did in the $U(1)$ case, we can pass to a KK-mode
expansion and calculate the effective interaction
amongst the KK-modes. Again, the KK-mode parity is now
locked to space-time parity and the CS term yields new
interactions, including 4-body amplitudes.  

As an example application of this formalism 
we turn presently to something slightly different
and give a simple derivation of the chiral constituent quark
model and the Wess-Zumino-Witten term \cite{wess,witten}
from a Yang-Mills theory in $D=5$.

\subsection{Application: The Wess-Zumino-Witten Term}
 
Consider the theory truncated on the zero-mode $A^5$ field.
All Stueckelberg gauge fields $B_\mu$ now become pure gauge fields:
\beq
B_\mu(x_\mu , y) = iU^\dagger(y) \partial_\mu U(y)
\eeq
and the field strength,  $G_{\mu\nu}$, is now zero.

The Wilson line, $U$, now contains only
the zero mode $A^5$ and the path-ordering is
no longer necessary because the $x^5$ wavefunction
is factorized from the flavor orientation:
\beq
U(y) = \exp \left(i\int_0^y A_5 \; dx^5 \right)
\eeq
We identify 
the Wilson line extending between
the two branes with a chiral field of mesons:
\beq
\label{chiral0}
W = \exp \left(i\int_0^R A_5 \; dx^5 \right) \equiv \exp (2i\tilde{\pi}/f_\pi) ,
\qquad \tilde{\pi} = \pi^a \lambda^a/2
\eeq
For the pseudoscalar octet, $f_\pi = 93$ MeV with this normalization.
We remark that this completely fixes the normalization of
our $\tilde{\pi}$ field. That is, the kinetic term,
$\Tr(G_{\mu 5}G^{\mu 5}) \sim \Tr(\partial\tilde{\pi})^2$
normalization determines $f_\pi$
in terms of $R$ and $\tilde{g}$, but the Wilson line completely
specifies the definition of $f_\pi$ in terms of the $A^5$ line integral
which is all we need presently (\ie, we can just assume the kinetic terms
are correctly normalized by the relationship between $\tilde{e}$
$R$ and $f_\pi$).

The $A^5$ wavefunction on
the orbifold in the $x^5$ coordinate is $\propto \sin(\pi x^5/R)$,
and using eq.(\ref{chiral0}) we can simply write:
\beq
U(y) = \exp (-2ih(y)\tilde{\pi}/f_\pi );
\qquad h(y) = \half\big( 1-\cos(\pi y/R) \big)
\eeq
We can actually use any monotonic wave-function $h(y)$
satisfying $h(0)=0 $ and $h(R) = 1$. 
Thus, expanding the pure gauge vector potential, we see:
\beq
\label{expand}
B_\mu(x_\mu , y) = - 2\times
\left(-\frac{1}{f_\pi}h(y)\partial_\mu \tilde{\pi} + i
\frac{1}{f^2_\pi}h(y)^2
[\tilde{\pi}, \partial_\mu \tilde{\pi}] + {\cal{O}}([\tilde{\pi},[ \tilde{\pi},
\partial \tilde{\pi} ] ]\right)
\eeq
Since we now have pure gauge configurations,
the only surviving term in
the CS-term of eq.(\ref{CS20}) is the 
$\Tr ((\partial_y B_\mu) B_\nu B_\rho B_\sigma)\epsilon^{\mu\nu\rho\sigma} $ 
term.
We note that 
$\Tr (B_\mu B_\nu B_\rho B_\sigma)\epsilon^{\mu\nu\rho\sigma} =0 $ 
by the cyclicity of the trace, which makes eq.(\ref{CS20}) easy
to evaluate when we
substitute the expression of  eq.(\ref{expand}), and it takes the form:
\bea
\label{CS21}
{\cal{L}}_{1} & = &
 \frac{8ic}{f_\pi^5}\epsilon^{\mu\nu\rho\sigma}\int d^4x \int_0^R dy\;
 (\partial_y h(y))\; h(y)^4\;
 \Tr \left[  [\tilde{\pi}, \partial_\mu \tilde{\pi}]
 \partial_\nu \tilde{\pi}\partial_\rho \tilde{\pi}\partial_\sigma \tilde{\pi}
 \right] + \; ...
\eea
where the ellipsis refers to higher powers in the $\tilde\pi$ field.

Since $h(0) = 0$  and $h(y)=1$ we see that the integral over $y$
is trivial -- the integrand is an exact differential and
the result doesn't depend upon the particular shape of the wavefunction
$h(y)$ ! For example, the zero-mode $A^5$ could have been a constant $x^5$ 
whence, $h(y) = y/R$, and we would obtain the same result.
The resulting expression is:
\bea
\label{CS21}
{\cal{L}}_{1} & = &
 \frac{2N_ci}{15f_\pi^5}\epsilon^{\mu\nu\rho\sigma}\int d^4x 
 \Tr \left[  \tilde{\pi} \partial_\mu \tilde{\pi}
 \partial_\nu \tilde{\pi}\partial_\rho \tilde{\pi}\partial_\sigma \tilde{\pi}
 \right]
\eea
Where we have removed the commutator in eq.(\ref{CS21}) using
cyclicity of the trace, and we have
substituted the coefficient, $c = 1/24\pi^2$. This
agrees with Witten's famous result \cite{witten}.
In ref.(\cite{hill4}) we develop the fully gauged WZW term 
from the present formalism.

We can develop a matching
to the QCD spectrum from a compactified
$D=5$ Yang-Mills theory by considering the boundary conditions
dual to an orbifold,
$\epsilon^{ABCD5}F_{AB}|_I = \epsilon^{ABCD5}F_{AB}|_{II}=0$,
equivalently, $F_{\mu\nu}|_I = F_{\mu\nu}|_{II}=0$.
This implies a ``flipped'' or
an ``anti-orbifold'' in which $A^5$ now
has a zero-mode with a flat wave-function in $x^5$ 
(and even basis functions on
the $[0,2R]$ interval), while $A_\mu$ is now odd, beginning
with a $J^P = 1^-$ massive mode, then a $J^P=1^+$ recurrence, etc. 
For an $SU(3)$ gauge group of flavor
this, remarkably, has the spectrum corresponding to the QCD mesons. 
The Wilson line, $\int dx^5 A_5$, over the zero-modes forms the octet
of {\em pseudoscalar} mesons (containing $\pi, K,\eta$), while the first
$A^5$ KK-modes correspond to the $0^+$ octet ($a^0$). The $A^\mu$ KK-mode tower
begins with the ($\rho$) vector meson octet, 
then the ($A^1$) axial-vector meson octet, etc.  We expect
the quantized coefficient of the CS term remains the
same in this model, hence the
WZW term remains the same. 
Here the CS term clearly exclusively becomes the
WZW term, because the boundary conditions 
$F_{\mu\nu}|_I = F_{\mu\nu}|_{II}=0$
prohibit any Stueckelberg field anomalies that might contribute
to the WZW term.

\section{Conclusions}

This paper has investigated the physics of the Chern-Simons term
in gauge theories in compactified extra dimensions.
The main thrust is that Chern-Simons terms must occur
in association with ``chiral delocalization,'' whereby anomalous
chiral fermions are placed in different locations in a $D=5$ bulk (\ie, ``split''
anomalies).
We view chiral delocalization as a compelling attribute of extra dimensional
theories, providing a rationale for the existence of flavor-chirality
(non-vectorlike representations)
as is seen in the standard model. We observe that even a vectorlike theory,
such as QED, must become chirally delocalized
when it is imbedded into $D=5$, to naturally protect the small
electron mass. The electron mass becomes a Wilson line connected the
chiral partners. In these cases the Chern-Simons term is inevitable.
We study two models, (1) chiral fermions on opposing branes and
(2) axions on branes.

For bulk propagating gauge fields,
The Chern-Simons term locks KK-mode parity to the parity of space-time. 
Indeed, KK-mode parity,
if it is present, is a spurious symmetry, independent of space-time parity.
The Chern-Simons terms blends these symmetries into a single
surviving parity. This is the exact
analogue of the pion parity for the Wess-Zumino-Witten term.

Let us summarize how the analysis procedes in general.  We begin 
in a $D=5$ gauge theory, compactified
in $0 \leq x^5 \leq R$, with chirally delocalized fermions
on the boundaries (branes).
The theory contains a bulk-filling Chern-Simons term.
The chiral fermions have a gauge invariant mass term that is bilocal,
$\sim\bar{\psi}_L(x,0)W\psi_R(x,R)+h.c.$, and
involves the Wilson line, $W=P\exp(i\int_0^R A_5dx^5)$ 
that spans the bulk. 
A general gauge transformation in the bulk produces anomalies on the
boundaries coming from the Chern-Simons term.  
Likewise, this gauge transformation produces anomalies, coming
from the fermions on the boundaries. These anomalies take the consistent form,
\ie, they are the direct result of the Feynman triangle loops for the
fermions, and have the identical same form 
as the anomalies from the CS term (see Appendix).
We demand that these anomalies cancel, and this fixes the coefficient of the 
CS term, generally to $c=1/24\pi^2$.

We now rewrite the CS term into a form that displays separately $A_5$ 
and $\partial_5$. We then perform a master gauge transformation that converts
$A_5\rightarrow B_5 = 0$, and $A_\mu \rightarrow B_\mu$. 
This also sets the Wilson line spanning the bulk between
the branes to unity. The massive components of 
the $B_\mu$ are now gauge invariant Stueckelberg fields,  having ``eaten''
their longitudinal degrees of freedom contained 
in the non-zero modes of $A_5$. 

Finally, we integrate out the fermions in the large $m$ limit. This
produces effective interactions (the log of the Dirac determinant) 
on the boundaries. The form of
this effective interaction is just Bardeen's counterterm \cite{bardeen}
that maps consistent anomalies into covariant ones.  We thus have an
expression for total action, ${S}_{full}$, the sum of $S_{CS}$, 
the Chern-Simons term, and
the boundary terms from the fermionic Dirac determinant,
summarized by the matrix elements of ${\cal{O}}_3$.
These are functionals of the Stueckelberg fields contained
in the mode expansion of  ${B}_\mu$. The new interactions involve
the ``structure constants,'' $\bar{c}_{nmk}$, for the different
KK-modes, $(n,m,k)$. We  demonstrate
that $S_{full}$  leads to 
new physical processes involving 3-body decay amplitudes
amongst KK-modes. These processes violate naive KK-mode parity, but
conserve the combined space-time and KK-mode parity. 
As an example of the formalism, we explicitly compute the decay 
widths for massive KK-modes into lighter KK-modes, plus the zero mode.

The Chern-Simons term coefficient was determined at the outset by cancelling
the anomalies of matter fields that are localized
on branes. We are lead to a study of the massless and
massive consistent anomalies of Weyl fermions. The relationship 
between the ``consistent'' and ``covariant'' anomalies 
involves a counterterm, (which is shown to be our boundary term,
term truncated on the zero-mode and first KK-mode of a $D=5$ theory).
The tower of KK-mode currents and their anomalies are determined
by the ``structure constants,'' $c_{nmk}$, that appear in 
the Chern-Simons term in a mode expansion.
The form of the covariant anomalies involves a miraculous identity
amongst the $c_{nmk}$.

We finally develop the non-abelian formalism, to the
point of computing the coefficient of the Chern-Simons term. 
This is similar to the $U(1)$
case, though we postpone the construction of the KK-mode
effective lagrangian and derivation of the structure constants.
As an example, however, we show how an $SU(N)$ Yang-Mills theory of
quark flavor can be compactified into a low energy $SU(N)_L\times SU(N)_R$
chiral langrangian, where the $A^{5n}$ gauge fields have become ``mesons.''
In this case the Chern-Simons term built of KK-mode of gauge fields, becomes
the Wess-Zumino-Witten term. We immediately obtain
Witten's coefficient for this WZW term, and the formalism will be developed
in a subsequent paper, yielding the fully gauged WZW term \cite{hill4}.

There is large number of theories to which these considerations apply. 
These include various incarnations
of Randall-Sundrum models, Little Higgs theories, and 
models of (anomaly) split fermion representations
in extra dimensions.  We further envision applications to
string theory, and AdS-CFT QCD as well. The WZW term of gravitation in
a split anomaly mode, \eg, in $D=6$ and $D=7$, would also be
an intriguing application.   
An interesting candidate for further study
is the standard model with left-handed fermions on brane $I$ and right-handed
fermions on brane $II$. An $SU(2)\times U(1)$ theory 
is a subgroup of $SU(3)$ and
does contain $d$-symbols, and a non-abelian CS term is present
to cancel the delocalized consistent matter anomalies. 
In this theory the Higgs mechanism has to yield the Wilson line connecting
the two branes, so the Higgs field is spread out in the bulk. 
This is remniscent of many ``Higgsless'' theories.  It is of interest
to explore the induced physics via the CS term in these models.
It would likewise be interesting to explore an axi-gluonic 
QCD that can be constructed by splitting left-color
onto $I$ and right-color onto $II$. Again, a quantized CS term will occur.
 
The background geometry has been taken flat in the present discussion. The
results may have limited sensitivity to the introduction of curvature. 
It would be interesting to reformulate the $U(1)$ model
in a Randall-Sundrum geometry to test the ideas. Moreover,
gravitational CS terms could be developed in a parallel manner.

Originally, in beginning this analysis, we had hoped
that the production
of single KK modes through the CS term would
be an available channel. For example, gluon fusion into
a colored axi-gluon would be a spectacular LHC signature,
or $e^+e^-\rightarrow \gamma+Z'$ might occur at a sufficiently
energetic linear collider.
The $U(1)$ case in the large $m_{electron}$ limit disallows such
processes by the cancellation of the triangle diagrams with
the CS-term for one KK plus two zero-mode vertices, as
in the coefficients $\bar{c}_{mnk}$. However, the axion
model shows that such processes can occur (though in association with photon
mass). We have not explictly
checked, however, that this process does 
not occur for massless zero-modes in Yang-Mills,
or considered processes such as ``associated KK mode production'' 
\eg, zero + zero $\rightarrow$ zero + KK.
Thus, the collider physics implications of the Chern-Simons term remain to
be investigated.

The $U(1)$ $D=5$ continuum bulk theory consisting 
only of gauge kinetic terms and the Chern-Simons term, with
matter restricted to the branes, is in itself
an interesting system. The bulk topological
theory is subject to some nonrenormalization constraints
and it would be interesting to study its loop
structure in $D=5$ further.

\appendix

\noindent\section{Axions on the Branes}

The simplest anomaly free model with the bulk filling
CS term involves the incorporation of ``left" ($\phi_L$)
and ``right" ($\phi_R$) axion fields confined to the 
respective boundary branes.
The  axions are coupled to the left- and right- anomalies,
and they can freely shift to absorb the 
induced boundary anomalies from the CS term 
under a gauge transformation.

The kinetic terms of the axions must 
be invariant under this shift, and this requires that they
couple longitudinally to the
left- and right-gauge fields on the respective branes. This 
further locks the axion shift
to the gauge transformation. In the $D=4$
effective theory, we then find that
one linear combination of the axions, $\phi^{(+)}$,
is eaten to become a longitudinal massive photon,
where:
\beq
\phi^{(+)} = \half (\phi_L +\phi_R) \qquad
\phi^{(-)} = \half (\phi_L-\phi_R)
\eeq
$\phi^{(-)}$ remains as a massless state in the spectrum, coupled
longitudinally to the pseudovector KK-modes. This is the physical axion.
The model exhibits the fact that the bulk CS term interactions
are physical, and yields $\bar{c}_{nmk} = c_{nmk}$. 

Consider QED in $D=5$ on an orbifold
with periodic domain $0 \leq x^5 \leq R$. The full
action of the theory is,
\beq
S=S_0 + S_{CS} + S_{branes}
\eeq
where the gauge field kinetic term action for the theory is:
\bea
S_0 & =& -\frac{1}{4\widetilde{e}^2}\int_0^R dy\int d^4 x\; F_{\mu\nu}F^{\mu\nu}
-\frac{1}{2\widetilde{e}^2}\int_0^R dy\int d^4 x \;
F_{\mu 5}F^{\mu 5} \\ \nonumber
\eea
The action $S_{CS}$ is defined in eq.(\ref{SCS}).
On brane $I$ at $y=0$ ($II$ at $y=R$),
we place an axion field $\phi_L(x^\mu)$ ($\phi_R(x^\mu)$), and we
define the action:
\bea
\label{axionanomaly}
S_{branes} & = &   \half \int_I d^4x \;
m^2(A_\mu(0,x_\mu) -\frac{1}{m}\partial_\mu \phi_L)^2
+ \half \int_{II} d^4x \;
m^2(A_\mu(R,x_\mu) -\frac{1}{m}\partial_\mu \phi_R)^2
\nonumber \\
& & + \frac{c}{4m}\int d^4x\; 
\phi_L\; \epsilon^{\mu\nu\rho\sigma}
F_{\mu\nu} F_{\rho\sigma}|_{I} 
-\frac{c}{4m}\int d^4x\;
\phi_R\; \epsilon^{\mu\nu\rho\sigma}
F_{\mu\nu} F_{\rho\sigma}|_{II} 
\eea
This construction converts our $U(1)$ gauge
theory into an effective chiral theory, \ie,
with distinct axion ``chiralities'' living on the two distinct branes. 
The  $m^2$ terms contain the axion kinetic
terms, as well as longitudinal couplings
to gauge fields on the respective branes, locking the $\phi$'s
to the gauge fields under gauge transformations. 
Thus a gauge transformation in the bulk:
\beq
\label{gtB}
A_A(x_\mu, y) \rightarrow A_A(x_\mu, y)
+ \partial_A \theta (x_\mu, y) ,
\eeq
implies, through the $m^2$ terms, a transformation
of the axions on the branes:
\beq
\label{axionshift}
\phi_L \rightarrow \phi_L + m\theta (x_\mu, 0);\qquad
\phi_R \rightarrow \phi_R + m \theta (x_\mu, R).
\eeq 
The CS term under this transformation 
generates anomalous terms on the branes,
as in eq.(\ref{test2}). These anomalies
are now cancelled by the matching
shifts in the axion fields coupled to
$F\widetilde{F}_I$ and $F\widetilde{F}_{II}$ as in eq.(\ref{axionanomaly})
and the theory is gauge invariant.

We can thus perform the Wilson line
gauge transformation of eq.(\ref{gt}) with impunity.
The brane action then becomes a functional
of the Stueckelberg fields $B_\mu$:
\bea
\label{axionaction2}
S_{branes} & = &  \half \int_I d^4x \;
m^2(B_\mu(0,x_\mu) -\frac{1}{m}\partial_\mu \phi_L)^2
+ \half \int_{II} d^4x \;
m^2(B_\mu(R,x_\mu) -\frac{1}{m}\partial_\mu \phi_R)^2
\nonumber \\
& & + \frac{c}{4m}\int_I d^4x\; 
\phi_L\; \epsilon^{\mu\nu\rho\sigma}
F_{B\mu\nu} F_{B\rho\sigma}|_{I} 
-\frac{c}{4m}\int_{II} d^4x\;
\phi_R\; \epsilon^{\mu\nu\rho\sigma}
F_{B\mu\nu} F_{B\rho\sigma}|_{II} 
\eea
where all gauge fields are now of the form
eq.(\ref{Stu}).

It is useful to write $B_\mu(x^\mu,y)$ in terms of
a zero mode, $A^0(x^\mu)$ which is independent of $y$,
and the non-zero mode $x^5$ parity components, 
$B^{+}_\mu(x^\mu,y)$ and $B^{-}_\mu(x^\mu,y)$,
which have
non-zero KK-mode momentum. We have
the $x^5$ parity assignments:
\bea
A^{0}_\mu(x^\mu,y) & = & A^{0}_\mu(x^\mu,R-y),
\qquad
B^{+}_\mu(x^\mu,y)  =  B^{+}_\mu(x^\mu,R-y),
\nonumber \\
B^{-}_\mu(x^\mu,y)  & = & -B^{-}_\mu(x^\mu,R-y).
\eea
Thus, on the branes we define:
\bea
B^{+}_\mu(x^\mu) & \equiv & B^{+}_\mu(x^\mu,0) = B^{+}_\mu(x^\mu,R),
\qquad
B^{-}_\mu(x^\mu)  \equiv  B^{-}_\mu(x^\mu,0) = -B^{-}_\mu(x^\mu,R).
\eea
The decomposition thus takes the form:
\beq
B_\mu(x^\mu,y) = A^0(x^\mu)+  
B^{+}_\mu(x^\mu,y) + B^{-}_\mu(x^\mu,y)
\eeq
We presently require only that
$A^0(x^\mu)$, $B^{+}_\mu(x^\mu,y)$ and $B^{-}_\mu(x^\mu,y)$
are orthogonal fields upon intergation over $y$ from $0$ to $R$.

After performing the gauge transformation of
eq.(\ref{gtB}), we have brought the field $A_5(x^\mu,y)=0$
everywhere throughout the bulk. There remains a residual
gauge transformation that we can do which maintains this
gauge condition,  which
redefines 
the zero mode (photon) field:
\beq
\label{ggt}
B^0_\mu(x^\mu) = A^0_\mu(x^\mu)+ \partial_\mu\theta(x^\mu)
\eeq
where:
\beq
\theta(x) = \frac{1}{{2}m}(\phi_L(x^\mu)+ \phi_R(x^\mu))
\equiv \frac{1}{m}\phi^{(+)}
\eeq
We do not shift the $\phi^{(+)}$ and $\phi^{(-)}$ fields
under this transformation. The bulk CS term, however, generates
the surface anomalies under this gauge transformation,
which cancel the $\phi^{(+)}$ anomalous coupling on the branes
(this can be explicitly checked by substituting eq.(\ref{ggt})
into the CS term  and
noting the integrand is an exact differential
in $y$).
This brings the photon field into the form of
a massive Stueckelberg field 
and $\phi^{(+)}$.
thus disappears from the action. The brane action 
now takes the form:
\bea
\label{axionaction3}
S_{branes} & = &  \int d^4x \;\left[ \half\left(\partial_\mu \phi^{(-)}\right)^2+
m^2\left(B^0_\mu(x^\mu)+  
B^{+}_\mu(x^\mu) \right)^2 
+ m^2 \left(B^{-}_\mu(x^\mu)\right)^2 - mB^{-}_\mu(x^\mu)\partial^\mu
\phi^{(-)}\right]
\nonumber \\
& & 
+ \frac{c}{m}\int d^4x\; 
\phi^{(-)}\; [
F_{B^0\mu\nu} \tilde{F}_{B^0}^{\mu\nu} + 2 F_{B^0\mu\nu} \tilde{F}_{B^+}^{\mu\nu}
+ F_{B^+\mu\nu} \tilde{F}_{B^+}^{\mu\nu}
+F_{B^-\mu\nu} \tilde{F}_{B^-}^{\mu\nu}].
\nonumber \\
\eea
This shows that the theory contains a residual massless
axion, $\phi^{(-)}$ coupled to the anomalies, and
longitudinally to the pseudovector KK-modes. This is
remniscent of the $\pi^0$. It is not eaten since
the KK-modes, $B^+$ amd $B^-$, acquire (large) masses
from the bulk
kinetic terms. There will also be a
mass term induced for the photon of leading order order:
$
\approx m^2 \left[B^0(x^\mu)\right]^2
$
with corrections due to mixing with the KK-modes of order 
$m^4/M_{KK}^2 + ...$ 

Let us now consider the orbifold compactification in flat space-time.
The mode expansion follows the identical form as in the 
text, eqs.(\ref{mode}).
The axion coupling to the KK-modes is obtained
from substituting the mode expansions into eq.(\ref{axionaction3}).
We define:
\beq
\tilde{m}^2 = \frac{2m^2\tilde{e}^2}{R}
\eeq
and we have the Stueckelberg field for
the massive photon with our zero mode normalization:
\beq
B^0_\mu = A_\mu^0
-\frac{1}{\sqrt{2}\tilde{m}}\partial_\mu \phi^{(+)}
\eeq

The full $D=4$ effective action, including the brane axion
component  of eq.(\ref{axionaction3}), together
with the bulk gauge field kinetic terms and CS term, becomes:
\bea
\label{axionaction4}
S_{axion} & = & 
 \int d^4x \left[ \half\left(\partial_\mu \phi^{(-)}\right)^2
  +
\tilde{m}^2\left(\frac{1}{\sqrt{2}}B^0+  
\sum_{n\;even}\!\! B^{n}_\mu \right)^2 
+\tilde{m}^2 \left(\sum_{n\;odd}B^{n}_\mu\right)^2 
\!\! - \tilde{m}B^{-}_\mu\partial^\mu
\phi^{(-)}\right]
\nonumber \\
& & 
\!\!\!\!\!\!\!\!\!\!\!
+\; \frac{c}{2\tilde{m}}\int d^4x\;\phi^{(-)}\;\epsilon^{\mu\nu\rho\sigma} 
\times 
\nonumber \\
& & 
\;\;\;\;\; \left[\;
e^2e' F^{B^0}_{\mu\nu} F^{B^0}_{\rho\sigma}  
+ 2ee'{}^2 F^{B^0}_{\mu\nu}\sum_{n \;even} F^{n}_{\rho\sigma}
 + e'{}^3\sum_{n,m\; even} F^{n}_{\mu\nu} F^{m}_{\rho\sigma}
+e'{}^3 \sum_{n,m\; odd} F^{n}_{\mu\nu} F^{m}_{\rho\sigma}\right]  
\nonumber \\
& &  
\!\!\!\!\!\!\!\!\!\!\! +
\int d^4x\; 
\left[
 \;+\frac{1}{12\pi^2} \sum_{nmk}e_n e_m e_kc_{nmk}B^n_\mu B^m_\nu 
\widetilde{F}^{k\mu\nu}
+\sum_{n}\left( -\frac{1}{4} F^n_{\mu\nu}F^{n\mu\nu}+ 
\half M_n^2 B_\mu^n B^{n\mu}\right) \right].
\nonumber \\ &&
\eea
This effective action is characterized by the presence of
the physical axion, $\pi^-$, the photon mass term, 
$\tilde{m}^2 B^0_\mu B^{0\mu}$ and the pure CS term structure constants,
$\bar{c}_{nmk} = c_{nmk}$.


\section{KK-mode Currents and Covariant Anomalies }

We derive the currents 
of the theory of eq.(\ref{effective2}) by variation of the full action wrt
$B^n_\mu$. The spinor currents are supplemented 
by current contributions from the Chern-Simons term: 
\bea
\tilde{J}^{n}_\mu & = & \frac{\delta S}{\delta B^{n\mu}} =
\bar{\psi}\gamma_\mu \psi|_{n\;even} 
+\bar{\psi}\gamma_\mu \gamma^5\psi|_{n\;odd} + J_{\mu}^{n\;CS}
\eea
where $J^{n\;CS}$ is the Chern-Simons current:
\bea
J_{mu}^{n\;CS} = \frac{\epsilon_{\mu\nu\rho\sigma}}{12\pi^2} 
\sum_{mk}\bigl[(c_{nmk}-c_{mnk}+ 
c_{kmn}-c_{mkn}) B^{m\nu }
\partial^\rho {B}^{k\sigma} \bigr] 
\eea
In what follows
we use the fermionic current 
consistent anomalies as computed in \cite{bardeen}
(Appendix C.4).

We can now  compute the current divergences.
The consistent anomalies can be
written in one compact formula for both  axial
and vector currents as (from eq.(\ref{bardeen11})): 
\bea
\label{an}
\partial^\mu J_\mu^n & = & \frac{1}{48\pi^2}
\sum_{mk}\left( 1-(-1)^{n+m+k}\right){F}^m_{\mu\nu} \widetilde{F}^{k\mu\nu}
\eea
The $J^n_\mu$  are the axial currents
for $n$ odd ($J^{(n\; odd)} = \bar{\psi}\gamma_\mu \gamma^5\psi $ )
and vector currents for  $n$ even 
($J^{(n\; even)} = \bar{\psi}\gamma_\mu\psi $ ).

Moreover, the divergence of the Chern-Simons current
takes the form:
\bea
\partial^\mu J_{\mu}^{n\;CS}  & = & 
\frac{1}{48\pi^2}\sum_{m,k}
(c_{nmk}-c_{mnk}+ c_{nkm}-c_{knm})
{F}^m_{\mu\nu} \widetilde{F}^{k\mu\nu}
\eea
where we have exploited the even symmetry under
interchange of indices of $k\leftrightarrow m$ 
to resymmetrize the summand.
Thus, the full current divergence takes the form
\bea
\partial^\mu \tilde{J}^{n}_\mu   & = & 
\frac{1}{24\pi^2}\sum_{m,k}
d_{nmk}
{F}^m_{\mu\nu} \widetilde{F}^{k\mu\nu}
\eea
where we have a remarkable identity:
\bea
\label{dd}
d_{nmk} & = &  \half \left[(1-(-1)^{n+m+k})+
(c_{nmk}-c_{mnk}+ c_{nkm}-c_{knm})
\right]
\nonumber \\
& = &  \frac{3}{2}[(-1)^{n+m+k}-1]\frac{n^2(k^2+m^2-n^2)}{(k+m-n)
(k+m+n)(k+n-m)(k-m-n)}
\nonumber \\
& = &  \frac{3}{2}c_{nmk}
\eea
where the $c_{nmk}$ are defined in eq.(\ref{ccc}).
Hence, we have the final result for the entire KK-tower
of currents:
\bea
\partial^\mu \tilde{J}^{n}_\mu   & = & 
\frac{1}{16\pi^2}\sum_{m,k}
c_{nmk}
{F}^m_{\mu\nu} \widetilde{F}^{k\mu\nu}
\eea
where $n$ even (odd) is a vector (axial vector) current.
The conservation of the electromagnetic
current is now manifest, since:
\beq
c_{0mk} = 0
\eeq
while the remaining $n\geq 0$ currents have anomalies.

In particular, as a check of this result,
if the theory contains only the $n=0$
photon, $\gamma$, and the $n=1$ axial vector meson, $B$, we then have 
a conserved vector electromagnetic current and we find the
divergence of the full axial vector current:
\bea
\label{vresult}
\partial^\mu J_{\mu}^{5} & = & 
\frac{1}{16\pi^2}\left(c_{100}
{F}_{\gamma\;\mu\nu} \widetilde{F}_{\gamma}^{k\mu\nu}
+ c_{111}
{F}_{B\;\mu\nu} \widetilde{F}_{B}^{k\mu\nu}
\right)
\nonumber \\ 
& = & 
\frac{1}{8\pi^2}
{F}_{\gamma\;\mu\nu} \widetilde{F}_{\gamma}^{\mu\nu}
+\frac{1}{24\pi^2}
{F}_{B\;\mu\nu} \widetilde{F}_{B}^{\mu\nu}
\eea
The latter expression is Bardeen's result for the axial
vector anomaly when he enforces a conserved vector current.
The CS term is playing the role of Bardeen's counterterm, which
brings the current divergences into the present form \cite{bardeen}.

The general issue of anomaly consistency is that
the full currents are sources of the gauge 
field equations of motion:
\bea
\partial^\mu F^0_{\mu\nu} & = & e\tilde{J}^0_\nu \nonumber \\
\partial^\mu F^n_{\mu\nu} + M_n^2B_\nu^n & = & e'\tilde{J}^n_\nu \qquad (a\neq 0)
\eea
Since $\partial^\mu\partial^\nu F^0_{\mu\nu}=0$ 
by antisymmetry of $F_{\mu\nu}$ we see that 
the full $\tilde{J}^0_\mu$ must be conserved, as indeed it is.
Since the axial vector
mesons have masses (and are Stueckelberg fields with imbedded
longitudinal components) there
is not an inconsistency here either, as the double divergence
implies:
\beq
\label{eom}
 M_n^2\;\partial^\nu B_\nu^n  =  \partial^\nu \tilde{J}^n_\nu =
 \makebox{(anomaly)}_n.
\eeq
This is just the equation of motion of the longitudinal modes,
the $A_5^n$ which are eaten by the $B_\mu^n$ fields.


\section{Triangle Diagrams }

The present Appendix is schematic and we only quote 
the main results.
A detailed description of these calculations
is available elsewhere \cite{hill3}.

The massless  calculation yields a result equivalent to
Bardeen's result for the consistent anomalies \cite{bardeen}. 
Bardeen performs, however, a
massive spinor loop calculation and quotes the anomalies 
in $V\pm A$ form.  All anomalies  we obtain presently
fully confirm Bardeen's result in both the massless and massive cases.  
We do, however,  see a slight subtlety in the form of the
pure current divergences expressed in the $V_L$ and $V_R$ forms
obtained in the massive case (the current divergence are 
not the full anomaly in
that case, since the $im\bar{\psi}\gamma^5\psi$ 
term must be subtracted. We also obtain the effective operator
description of the these currents used in the text.

\subsection{Single Massless Weyl Spinor}

Consider the action:
\beq
S_L = \int d^4x\; \bar{\psi}_L(i\slash{\partial} +\slash{V}_L)\psi_L
\eeq
where the guage fields:
\beq
V_{L\mu} = B^a_{L\mu} +B^b_{L\mu} + B^c_{L\mu}  
\eeq
couple to the current:
\beq
J_{L\mu} = \bar{\psi}_L \gamma_\mu \psi_L
\eeq
and the components of
$V_L$ have the respective masses $M^{a}$, $M^{b}$, $M^{c}$. 
We are compute the triangle loop with
three distinct external fields, $B^a$, $B^b$ and $B^c$, these
can be alternatively viewed as distinct momentum components of the
single field $V$. If all three fields were identical (exact
Bose invariance) the amplitude would vanish, since
it would involve an operator $VVdV$ which is zero. 
It is the external momentum differences 
or flavor indices that distinguish these fields
and allow non-zero operators such as $[B] \sim B^a B^b dB^c$ 
and $[B] \sim B^a B^c dB^b$, \etc. 
In the massless Weyl fermion case of interest presently, we compute
in a limit $M^a >> M^b \sim M^a \sim 0$. We can view 
this as an operator product expansion
of the triangle diagrams in which the internal lines carrying
$p^2 = M_a^2$ are treated as a short-distance expansion.

\begin{figure}[t]  
\vspace{5cm}  
\includegraphics{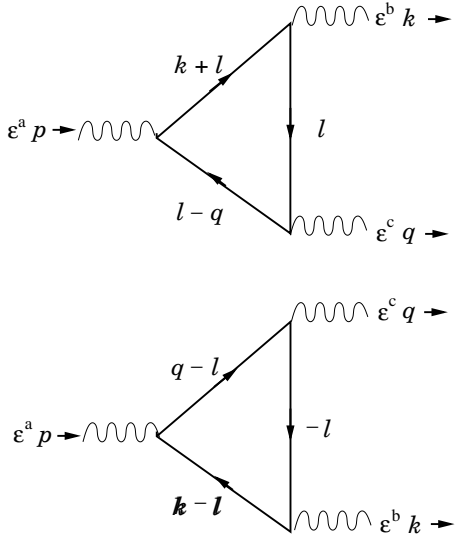}  
\vspace{4.5cm}  
\caption[]{
\addtolength{\baselineskip}{-.3\baselineskip}  
Bose symmetric triangle diagrams 
for $B^a(p)\rightarrow B^b(k)+B^c(q)$
The external lines are on mass-shell, $p^2=M_a^2 $, $k^2=M_b^2$
and $q^2=M_c^2$. The respective polarizations are $\epsilon^a_\mu$,
$\epsilon^b_\mu$ and $\epsilon^c_\mu$. The internal momentum
routing and integration momenta are chosen so that 
both diagrams have a common
denominator. }  
\label{dirac2}  
\end{figure}

With the particular choice of momentum
routing in the figure, we have the following expression for the sum
of the triangle diagram and its Bose symmetric counterpart,
which have a common denominator:
\bea
 T & = & (-1)(i)^3 (i)^3 \int \frac{d^{4}\ell}{(2\pi)^{4}}
\frac{N_1+N_2}{D} \nonumber \\
N_1 & = & \Tr[\slash{\epsilon}_a L
(\slash{\ell}-\slash{q})
\slash{\epsilon}_c L(\slash{\ell})
\slash{\epsilon}_b L(\slash{\ell}+\slash{k})]
\nonumber \\
N_2 & = & -\Tr[\slash{\epsilon}_a L
(\slash{\ell}+\slash{k})
\slash{\epsilon}_b L(\slash{\ell})
\slash{\epsilon}_c L(\slash{\ell}-\slash{{q}})]
\nonumber \\
D &  = &
(\ell+k)^2(\ell^2)(\ell-q)^2
\eea
where,
\beq
p=k+q, \qquad L =\half(1-\gamma^5), \qquad R =\half(1+\gamma^5).
\eeq
The overall sign contains: $\times(i)^3$ (vertices; note
that our vector potentials have the opposite sign to the
conventions of Bjorken and Drell, hence flipping the vertex
rule from $-i\gamma_\mu \rightarrow +i\gamma_\mu$)), $\times (i)^3$
(propagators), $\times (-1)$ (Fermi statistics).
In $N_2$ we've factored out an overall minus sign.
Note that one must use extreme care to write the 
given correct cyclic ordering of
the factors that make up the numerator, relative to
the momentum routing signs \cite{hill2}. This affects the overall sign of 
the triangle loop with three gauge vertices (but is has
no effect upon the $im\bar{\psi}\gamma^5\psi $ loop computed in the
massive case).

We unify the denominator using:
\beq
\frac{1}{ABC} = 2\int_0^1dy\int_0^y dz \frac{1}{(Az + B(y-z)+C(1-y))^3}
\eeq
The unified denominator becomes:
\beq
\frac{1}{D} = 2\int_0^1dy\int_0^y dz 
\frac{1}{( \ell^2 +2\ell\cdot(zk-(1-y)q)+zk^2 +(1-y)q^2  )^3}
\eeq
Shifting the loop momentum to a symmetric
integration momenta, $\overline{\ell}$:
\bea
 \ell & = & \overline{\ell} - zk+(1-y)q
\eea
the unified denominator becomes:
\bea
&  & (\overline{\ell}^2 +z(1-z)k^2 +y(1-y)q^2 
+2k\cdot qz(1-y))
\eea
(see the comment on shifting momenta below).

We define the following vertex tensors :
\bea
\label{match}
A & = & \epsilon_{\mu\nu\rho\sigma}{\epsilon}^\mu_a{\epsilon}^\nu_b
{\epsilon}^\rho_c k^\sigma \qquad \longleftrightarrow \qquad -i \bra{b,k;c,q } 
\epsilon_{\mu\nu\rho\sigma}B^{a\mu} B^{c\nu}\partial^\rho B^{b\sigma}\ket{
a,p }
\nonumber \\
B & = & \epsilon_{\mu\nu\rho\sigma}{\epsilon}^\mu_a{\epsilon}^\nu_b
{\epsilon}^\rho_c q^\sigma  \qquad \longleftrightarrow \qquad 
i \bra{ b,k;c,q  } \epsilon_{\mu\nu\rho\sigma}B^{a\mu}B^{b\nu}
\partial^\rho  B^{c\sigma} \ket{a,p }
\nonumber \\
C & = & \epsilon_{\mu\nu\rho\sigma}{\epsilon}^\mu_a{\epsilon}^\nu_b
{k}^\rho q^\sigma  \qquad \longleftrightarrow  \qquad
\half \bra{ b,k  } F_{\mu\nu}^a \tilde{F}^{b\;\mu\nu} \ket{a,p }
\nonumber \\
D & = & \epsilon_{\mu\nu\rho\sigma}{\epsilon}^\mu_a{\epsilon}^\nu_c
{k}^\rho q^\sigma \qquad \longleftrightarrow  \qquad
-\half \bra{ c,q  } F_{\mu\nu}^a \tilde{F}^{c\;\mu\nu} \ket{a,p }
\nonumber \\
E & = & \epsilon_{\mu\nu\rho\sigma}{\epsilon}^\mu_b{\epsilon}^\nu_c
{k}^\rho q^\sigma  \qquad \longleftrightarrow  \qquad
\half \bra{ b,k;c,q  } F_{\mu\nu}^b \tilde{F}^{c\;\mu\nu} \ket{ 0}
\eea
where we have indicated the corresponding 
operator matrix elements, $\bra{out }{\cal{O}}\ket{in} $
and note:
\beq
\tilde{F}_{\mu\nu} = \half \epsilon_{\mu\nu\rho\sigma}F^{\rho\sigma}
\eeq
is the standard definition of the dual field strength.

It is a well known ambiguity of the triangle loops in momentum
space that shifting
loop momenta can lead to residual terms, owing to the superficial
linear divergence. Such terms can only be of the
form $\propto [A] - [B]$, and they would yield
an anomaly that does not respect Bose symmetry in 
the three $a$, $b$ and $c$ channels. For example,
setting $\epsilon_a \rightarrow p_\mu$
we obtain $\rightarrow -2[E] \sim -2F_L\tilde{F}_L $, while
setting $\epsilon_c\rightarrow -q$ we obtain 
$\rightarrow +[C] \sim +F_L\tilde{F}_L$
Imposing this symmetry on the triangle loops is a luxury we
have only in the massless Weyl case, since we do not
have to subtract 
$\bra{b,c}im\bar{\psi}_L\gamma^5\psi_R + h.c.\ket{0}$
to obtain the anomaly. 
Imposing Bose symmetry on the anomaly as
constraint on the calculation removes the surface term ambiguity.
It turns out, however, for the particular
momentum routing we have chosen the result for the anomaly
is fully and nontrivially Bose symmetric, 
thus there is no surface term. Moreover,
even the superficial log divergence (also $\propto [A] - [B]$)
is cancelled, as we see below,
and the triangle loops are UV finite.

We now compute the triangle loops.  Since we are mainly
interested in a heavy KK mode decaying to low mass 
KK-modes, kinematically we have:
\beq
p = k+q \qquad M_b^2 = k^2 \approx 0 \qquad M_c^2 = q^2 \approx 0 \qquad
M_a^2 \approx 2k\cdot q 
\eeq
Hence the large $M_a^2$ limit corresponds to a symmetrical expansion
in $k^2/2k\cdot q $ and $q^2/2k\cdot q$.

For the large electron mass limit, $M_a^2 <<m^2$, 
we define the loop integrals with the usual Wick rotation
on the loop energy 
$\bar\ell_0$ and a Euclidean momentum space cut-of $\Lambda^2$:
\bea
\int \frac{d^4\bar\ell}{(2\pi)^4}\frac{(1,\;\bar\ell^2)}{(\bar\ell^2 -m^2+i\epsilon)^3}
& = & \left[ \frac{-i}{16\pi^2}\left(  \frac{1}{2m^2 }\right), 
\frac{i}{16\pi^2}\left[\ln\left(\frac{\Lambda^2}{m^2}\right) 
- \frac{3}{2}\right]\right]
\nonumber \\
\int \frac{d^4\bar\ell}{(2\pi)^4}\frac{(1,\;\bar\ell^2)}{(\bar\ell^2 -m^2+i\epsilon)^4}
& = &
\left[ \frac{i}{16\pi^2}\left(  \frac{1}{6(m^2)^2 }\right),
 \frac{-i}{16\pi^2}\left(  \frac{1}{3(m^2)^2 }\right)\right]
\eea
The familiar Wick rotation is a counterclockwise rotation of the contour
of the  $\bar\ell_0$ integral in the complex plane. The rotation is
clockwise to avoid the poles at $\pm\sqrt{\vec{l}^2+m^2}\mp i\epsilon$,
in the resulting Euclidean integral. For us, $m^2$ is actually
$m^2-z(1-y)M_a^2$ from the unified denominator.
In the case of $M_a^2 >>m^2$  the 
the $\vec{\ell}^2$ integrals develop
a cut structure. Nonetheless, 
the results are analytic functions of large
$m^2$, and in the limit $M_a^2 >> m^2$
we can simply replace $m^2\rightarrow -z(1-y)M_a^2$. In the
massless spinor case the  $zy$ integrals now acquire
infrared singularities, but massive case and the 
the anomaly are infrared finite.  

The diagrams yield a superficial 
log divergence. With $k^2=q^2=0$ (the $\alpha_0$ 
contribution) and upon doing the denominator unification integrals,
this yields a finite result:
\bea
\label{an1}
T_{\ln(\Lambda^2)} 
& = &
 \frac{2i}{16\pi^2}\;\int_0^1dy\int_0^y dz\;
(2i)\big[(1-3z)A + (2-3y)B \big]\left[\ln\left(\frac{\Lambda^2}{-z(1-y)M_a^2 }\right) 
-\frac{3}{2} \right]
\nonumber \\
& = &
-\frac{1}{24\pi^2}[A]+\frac{1}{24\pi^2}[B]
\eea
Note that $T_{\ln(\Lambda^2)}$ is Bose symmetric under interchange of
the photon and $b$-KK mode (let $q\leftrightarrow k$, hence $M_b^2\rightarrow
0$,
$\epsilon_a\leftrightarrow
\epsilon_b$, and note $A\leftrightarrow B$). However,
as mentioned above, this term cannot be the full result, because
its anomaly is not Bose symmetric, and we indeed must keep
residual finite terms. 

Note, as a check on
the large $m^2$ case, that if the argument of
the log, ${-\Lambda^2}/{z(1-y)M_a^2 }$ is replaced by ${\Lambda^2}/{m^2}$
then $T_{\ln(\Lambda^2)} = 0$. The result is finite because:
\bea
\label{an2}
\int_0^1dy\int_0^y dz\;
\big[(1-3z)A + (2-3y)B \big]
& = & 0
\eea
This furthermore implies that the imaginary part of
the expression is vanishing.

Combining all of
terms of the Feynman diagrams
yields the following full result for the triangle
diagrams (see \cite{hill2} for the detailed calculation):
\bea
T & = & 
-\frac{1}{12\pi^2}[A] + \frac{1}{12\pi^2}[B]
\nonumber \\
& & +\frac{1}{4\pi^2}[Ck\cdot \epsilon_c  + Ak\cdot q]\frac{I_b}{M_a^2}
\nonumber \\
& & -\frac{1}{4\pi^2}[C q\cdot \epsilon_c  + Aq^2]\frac{I_c}{M_a^2}
\nonumber  \\
& & +\frac{1}{4\pi^2}[D\epsilon_b\cdot k  + B k^2 ]\frac{I_b}{M_a^2}
\nonumber  \\
& & -\frac{1}{4\pi^2}[D\epsilon_b\cdot q  + B k\cdot q ]\frac{I_c}{M_a^2}
\nonumber  \\
& &
-\frac{1}{4\pi^2\; M_a^2}[E](\epsilon_a\cdot k \; I'_b + \epsilon_a\cdot q \; I'_c)
+ {\cal{O}}(q^2,k^2).
\eea
The integrals
$I_i$
and $I'_i$ are infrared divergent in our expansion. The result
is manifestly Bose symmetric only if we perform the unification
integrals, $I_i$ and $I'_i$ with 
a Bose symmetric IR cut-off. For a particular choice of small 
IR cut-offs
$x_i$, the leading log divergent terms are:
\bea
I_b & = & \int_0^{1-x_b}\int_0^y dz\;dy\;\frac{z(z-y)}{z(1-y)} 
= \half \ln(x_b) + k_b
\nonumber \\
I_c & = & \int_0^1\int_x^y dz\;dy\;\frac{(1-y)(z-y)}{z(1-y)} 
=\half \ln(x_c) + k_c
\nonumber \\
I'_{b}& = &\int_0^{1-x}\int_0^y dz\;dy\;\frac{2z-zy-z^2}{z(1-y)} = - \half \ln(x_b)
+k'_b
\nonumber \\
I'_{c}& = & \int_0^1\int_x^y dz\;dy\;\frac{z+y-zy-y^2}{z(1-y)} = - \half \ln(x_c)
+k'_c
\eea
Note that $I_b=I_c$ and $I'_b=I'_c$ 
if $x_b = x_c$ and Bose symmetry is maintained.
The physical cutoffs are of order 
$x_b \sim M_b^2/M_a^2$ and $x_c \sim M_c^2/M_a^2$,
and  $k$ and $k'$ are indeterminate.
This can be replaced with a more physical procedure by
resumming $k^2$ and $q^2$ into the denominators.
The logarithmic IR singularities in, \eg, the $ q^2 =0$
limit are presumably cancelled by collinear $\bar{\psi}\psi $ 
propagation in the $B^c\rightarrow B^b+ \bar{\psi}+\psi$
process, where $\bar{\psi}+\psi$ rescatter into a photon.

Note, however, a final lemma that is relevant to the anomaly:
\beq
\label{sumI}
I_a + I_b + I'_b + I'_c = 2
\eeq
which is an infra-red ``safe'' quantity.

\subsection{Massless Weyl Spinor Anomaly}

The amplitude we have just computed is:
\beq
T 
= \bra{b,c}T ...\; 
i\int d^4 x\; \exp(-ip\cdot x)\epsilon^{a}_{\mu}\bar{\psi}\gamma ^\mu
\psi_L \; ... \ket{0}
\eeq
On the other hand, the amplitude we want is the matrix
element of the current divergence:
\bea
W  & = & \bra{b,c}T ... 
\int d^4 x\; \exp(-ip\cdot x)\partial_{\mu}\bar{\psi}\gamma ^\mu
\psi_L \; ... \ket{0} \nonumber \\
& = & \bra{b,c}T ... 
\int d^4 x\; (-\partial_{\mu}\exp(-ip\cdot x)) \bar{\psi}\gamma ^\mu \psi_L
\; ... \ket{0}
\nonumber \\
& = & \bra{b,c}T ... 
\int d^4 x\; \exp(-ip\cdot x) ip_\mu \bar{\psi}\gamma ^\mu \psi_L
\; ... \ket{0}
\eea
We thus obtain $W$ from $T$ by the replacement:
\beq
W = T(\epsilon^{a}_{\mu}\rightarrow p_\mu)
\eeq
Under this substitution we have:
$[A]\rightarrow -[E]$, $[B]\rightarrow [E]$,
$[C]\rightarrow 0$, $[D]\rightarrow 0$,  
and  $\epsilon_a\cdot k \rightarrow k\cdot q$, 
$\epsilon_a\cdot q \rightarrow k\cdot q$.

We thus obtain:
\bea
T & \rightarrow  & W \; = \; 
\frac{1}{12\pi^2}[E] + \frac{1}{12\pi^2}[E] -
\frac{1}{8\pi^2}(I_a + I_b + I'_b + I'_c)[E]
\nonumber \\
& = & -\frac{1}{12\pi^2}[E]\qquad 
\eea
where we use eq.(\ref{sumI}).
The result is infra red non singular. 
Note that we have the operator correspondence
$[E]\rightarrow (1/4)F\tilde{F}$ from eq.(\ref{match}).
Our result for the anomaly thus
corresponds to the operator equation:
\beq
\label{resultL}
\partial^\mu \bar{\psi}\gamma_\mu\psi_L =
-\frac{1}{48\pi^2}F_{L\mu\nu}\tilde{F}^{\mu\nu}_L 
\eeq
It is trivial to infer the right-handed current
anomaly by flipping signs:
\beq
\partial^\mu \bar{\psi}\gamma_\mu\psi_R =
\frac{1}{48\pi^2}F_{R\mu\nu}\tilde{F}^{\mu\nu}_R 
\eeq
Defining $L=V-A$ and $R=V+A$ we can write this in the form:
\bea
  \partial^\mu \bar{\psi}\gamma_\mu \psi & = &
\frac{1}{12\pi^2}F_{V\mu\nu}\tilde{F}_A^{\mu\nu} 
\nonumber \\
\partial^\mu \bar{\psi}\gamma_\mu\gamma^5\psi & = &
\frac{1}{24\pi^2}(F_{V\mu\nu}\tilde{F}_V^{\mu\nu} 
+F_{A\mu\nu}\tilde{F}_A^{\mu\nu} )
\eea
which agrees 
with Bardeen's result for the left-right symmetric
anomaly \cite{bardeen}.

As a further check on the calculation, 
we can also examine the anomaly in the $B^c$ current, 
by letting 
$\epsilon_c\rightarrow -q$ (the minus
sign occurs since $B^c$ is outgoing), and we take the $c$ field to be
 on-shell and massless, \ie.
set $q^2=0$. 
 Whence $[A] =[C]$ and $[B]=[D]=[E] =0$:
\bea
T & \rightarrow &  -\frac{1}{12\pi^2} [C] 
\eea
Using eq.(\ref{match}) this corresponds to eq.(\ref{resultL}),
consistent with the $a$ channel result. 
Likewise, we can check the $B^b$ channel, and 
verify the same result.
We can, furthermore, check the
off-shell gauge invariance for $c$ identified
with a photon, and $M_c^2 =0$,
setting $\epsilon_c\rightarrow -q$ and examining
the  ${\cal{O}}(q^2)$ terms. These are found to cancel \cite{hill2}.
This implies that the only non-gauge invariant 
part of the amplitude is the
anomaly.

\noindent
\subsection{Finite Electron Mass}

We now turn to the case
of a finite, and large electron mass, where ``large''
means in comparison to external momenta and masses.
We carry out the analysis of the loops in
the presence of the full electron mass term, with the couplings
\beq
\int d^4x\; \left( \bar{\psi}_L(i\slash{\partial} +\slash{V}_L)\psi_L
+\bar{\psi}_R(i\slash{\partial} +\slash{V}_R)\psi_R
-m(\bar{\psi}_L\psi_R
+ h.c.)\right)
\eeq
where we take separate $L$ and $R$ fields,
$B_\mu^{a\;L,R}$:
\beq
V_{L\mu} = B^{aL}_\mu + B^{bL}_\mu + B^{cL}_\mu \qquad \qquad
V_{R\mu} = B^{aR}_\mu + B^{bR}_\mu + B^{cR}_\mu  
\eeq
Note that in comparison to the KK-mode normalizations used
in the text we have:
\beq
B^n_{L\mu} = (-1)^n B^n_{\mu} \qquad \qquad
B^n_{R\mu} = B^n_{\mu} 
\eeq 
We will implement this relationship subsequently, but presently
we work in the independent and generic  $V_L$, $V_R$ basis.

We presently adopt an obvious generalized notation for vertices, \eg,
\bea
A^{LRL} & = & \epsilon_{\mu\nu\rho\sigma}{\epsilon}^{L\mu}_a{\epsilon}^{R\nu}_b
{\epsilon}^{L\rho}_c k^\sigma \;,
\qquad\qquad 
A^{LRR}  =  \epsilon_{\mu\nu\rho\sigma}{\epsilon}^{L\mu}_a{\epsilon}^{R\nu}_b
{\epsilon}^{R\rho}_c k^\sigma \qquad   .\;.\;. 
\nonumber \\
C^{LR} & = & \epsilon_{\mu\nu\rho\sigma}{\epsilon}^{L\mu}_a{\epsilon}^{R\nu}_b
{k}^\rho q^\sigma  \qquad   .\;.\;. 
\eea
and so forth. 

We have just computed the $LLL$ 
($RRR$) loops arising from the pure 
massless $\psi_L$  ($\psi_R$). In the case of  a massive
electron 
the $LLL$ ($RRR$) loops have the same
numerator structure, but the  denominator now 
contains  electron mass terms:
\bea
D &  = &
[(\ell+k)-m^2][(\ell^2)^2-m^2][(\ell-q)^2-m^2]
\eea
This causes all of the previously computed $LLL$ 
($RRR$) terms to become suppressed
in the large $m^2$ limit. For example, the $\alpha_0$ term
previously computed for $m^2=0$
now becomes:
\bea
\label{an3}
T_{\ln(\Lambda^2)}
& = &
 -\frac{1}{4\pi^2}\;\int_0^1dy\int_0^y dz\;
\big[(1-3z)A + (2-3y)B \big]\left[\ln\left(\frac{\Lambda^2}{m^2-z(1-y)M_a^2}\right) 
-\frac{3}{2} \right]
\nonumber \\
& \longrightarrow &
-\frac{M_a^2}{480\pi^2 m^2 }\left([A] - [B]\right),
\eea
and now vanishes in the large $m^2$ limit.
All of the new terms of interest in the massive electron
case arise from the numerator terms containing mass insertions.
This represent mixing from $\psi_L$ to the  $\psi_R$,
and thus generates new vertices, such as $[A]^{LRL}$, \etc.  

We compute the triangle loops with a
single pure left-handed $\epsilon^{aL}_\mu \gamma^\mu L$ vertex, carrying
in momentum $p$, and again noting the 
the cyclic order in which numerator terms are written, we have:
\bea
 T_L & = & (-1)(i)^3(i)^3\int \frac{d^{4}\ell}{(2\pi)^{4}}
\frac{N_1+N_2}{D} \nonumber \\
N_1 & = &
\Tr[\slash{\epsilon}_a L(\slash{\ell}-\slash{q}+m)
(\slash{\epsilon}^L_c L+\slash{\epsilon}^R_c R) (\slash{\ell}+m)
(\slash{\epsilon}^L_b L+\slash{\epsilon}^R_b R) (\slash{\ell}+\slash{{k}}+m)]
\nonumber \\
N_2 & = &  -\Tr[\slash{\epsilon}_a L(\slash{\ell}+\slash{k}-m)
(\slash{\epsilon}^L_b L+\slash{\epsilon}^R_b R)(\slash{\ell}-m)
(\slash{\epsilon}^L_c L+\slash{\epsilon}^R_c R)(\slash{\ell}+\slash{q}-m)]
\nonumber \\
D &  = &
[(\ell+k)^2-m^2][\ell^2-m^2][(\ell-q)^2-m^2]
\eea
Note the sign flips in the momentum and $m$ terms in $N_1$
and momenta in $N_2$, a consequence
of having factored out an overall minus sign.

We obtain the result (full details are available in \cite{hill2}):
\bea
 T_L & = & 
 (-4im^2)\times 2\int_0^1 dy \int_0^y dz\; \int \frac{d^{4}\bar\ell}{(2\pi)^{4}}
\nonumber \\ & & \left( \frac{-z[A]^{LRL} - y[B]^{LRL}
+(1-z)[A]^{LLR} +(1-y)[B]^{LLR}
+z[A]^{LRR} - (1-y)[B]^{LRR}}{
(\bar\ell^2 +z(1-z)k^2 +y(1-y)q^2 
+2k\cdot qz(1-y) - m^2  )^3} \right)
\nonumber \\
\eea
This result is negligible in the limit
$k^2, \; 2k\cdot q,\; q^2 >> m^2$. 
However, in the limit of large $m^2$
the result reduces to:
\bea
\label{TL}
 T_L & = & \frac{1}{24\pi^2}(
 [A]^{LRL} + 2[B]^{LRL}-2[A]^{LLR} -[B]^{LLR} -[A]^{LRR} + [B]^{LRR})
\eea
From this we can easily infer the result for a computation
of the triangle loops with a single pure  $\slash{\epsilon}_\mu^{aR}
R$
(right-handed) vertex:
\bea
 T_R & = & -\frac{1}{24\pi^2}(
 [A]^{RLR} + 2[B]^{RLR}-2[A]^{RRL} -[B]^{RRL} -[A]^{RLL} + [B]^{RLL})
\eea
Combining these we have:
\bea
 T_{L} +T_{R} & = & \frac{1}{24\pi^2}([A]^{LRL} + 2[B]^{LRL}
 -2[A]^{LLR} -[B]^{LLR} -[A]^{LRR} + [B]^{LRR}\nonumber \\
& & -[A]^{RLR} - 2[B]^{RLR}
+ 2[A]^{RRL} +[B]^{RRL} +[A]^{RLL} - [B]^{RLL})
\eea

For  KK-mode $B_\mu^n$ we have an $x^5$ wave-function
parity of $(-1)^{n}$, and $B_{\mu L}^n = (-1)^{n}B_{\mu R}^n 
= B_{\mu}^n$. 
The KK-modes are normalized so that an axial vector
(odd $n$) couples to $\bar{\psi}\gamma_\mu \gamma^5\psi$ with positive sign.
Thus, we can write:
\bea
 T_{L} +T_{R} & = & \frac{1}{24\pi^2}((-1)^{a+c}([A] + 2[B])
 -(-1)^{a+b}(2[A] +[B]) -(-1)^{a}([A] - [B])\nonumber \\
& & -(-1)^{b}([A] + 2[B])
+ (-1)^{c}(2[A] +[B]) +(-1)^{b+c}([A] - [B])
\eea
This can be put into a compact final expression:
\bea
\label{final}
 T_{L} +T_{R} & = & \frac{1}{12\pi^2}(f_{abc}[A]^{} + g_{abc}[B])
\eea
where:
\bea
f_{abc} & = & \half((-1)^{a+c} -2(-1)^{a+b} -(-1)^{a} 
-(-1)^{b} +2(-1)^{c} +(-1)^{b+c}) \nonumber \\
g_{abc} & = & \half(2(-1)^{a+c}-(-1)^{a+b} +(-1)^{a} 
-2(-1)^{b} +(-1)^{c} -(-1)^{b+c}).
\eea
Note that  if $a + b + c $ is even, then $f = g =0$, which is
the condition that a transition cannot occur!
But, of course, 
the {\em condition that a transition can occur} is $a + b + c $ odd.
When $a + b + c $ is odd, we can therefore write:
\bea
\label{final2}
f_{abc} & = & -(-1)^{a}-(-1)^{b} +2(-1)^{c} \nonumber \\
g_{abc} & = & (-1)^{a}-2(-1)^{b} +(-1)^{c} 
\eea
Under $b\leftrightarrow c$ we have $A\leftrightarrow -B$ and
thus $g_{abc}\leftrightarrow -f_{acb}$, which confirms Bose symmetry.
Under the exchange
$a\leftrightarrow b$ we have $B\rightarrow -B$ and
$A\rightarrow A+B$ (since the $k$ in the $A$ vertex now becomes $-k-q$
with the sign flip,  since $a$ is incoming momentum $k+q$). 
Thus the vertex becomes:
\beq
T_{L} +T_{R}\rightarrow  \frac{1}{12\pi^2}(f_{bac}[A]^{} + (f_{bac}-g_{bac})[B])
\eeq
and we immediately verify that $f_{bac} = f_{abc}$
and $f_{bac}-g_{bac} =g_{abc}$.
The amplitude is seen to be fully Bose symmetric (we leave the
verification of $a\leftrightarrow c$ Bose symmetry to the reader).

The vertex calculation can be represented by an operator of
the form:
\beq
\label{triop}
{\cal{O}}_3 =  -\frac{1}{12\pi^2}\epsilon^{\mu\nu\rho\sigma} 
\sum_{nmk}a_{nmk}B^n_\mu B^m_\nu\partial_\rho B_\sigma^k 
\eeq
where:
\beq
a_{nmk} = \half [1 - (-1)^{n+m+k}](-1)^{m+k}
\eeq
For the process $a \rightarrow b + c$ 
the matrix element of ${\cal{O}}$ takes the form (we've multiplied
by $+i$ from $e^{iS}$):
\bea
\label{an10}
i\bra{a}{\cal{O}}\ket{b,c} & = & \frac{1}{12\pi^2} \big[
(-a_{abc}+a_{bac}+a_{bca}-a_{cba})[B]
+ (a_{acb}-a_{cab}+ a_{bca}-a_{cba})[A]
\big]
\nonumber \\
\eea
and we see that (for $a + b + c$ odd):
\bea
-a_{abc}+a_{bac}+a_{bca}-a_{cba} & = & g_{abc} \nonumber \\
a_{acb}-a_{cab}+ a_{bca}-a_{cba} & = & f_{abc}
\eea

\subsection{Massive Left-Right Symmetric Anomaly}

The current divergence, 
$\partial_\mu\bar{\psi}\gamma_\mu\psi_L$, is obtained by
the replacement $\epsilon_\mu \rightarrow p_\mu$ in $T_L$.
We thus have that
$A \rightarrow -E$ and $B\rightarrow E$ and we make use of
the operator correspondence eq.(\ref{match}):
\bea
\label{div1L}
\partial_\mu\bar{\psi}\gamma_\mu\psi_L & = & \frac{1}{48\pi^2}(
F^L_{\mu\nu}\tilde{F}^R_{\mu\nu}+F^R_{\mu\nu}\tilde{F}^R_{\mu\nu})
\eea
\bea
\label{div1R}
\partial_\mu\bar{\psi}\gamma_\mu\psi_R & = & -\frac{1}{48\pi^2}(
F^L_{\mu\nu}\tilde{F}^R_{\mu\nu}+F^L_{\mu\nu}\tilde{F}^L_{\mu\nu})
\eea
We emphasize that this result is {\em not the anomaly}.
To extract the anomaly, we note that the equations of motion
yield the divergences of the spinor currents:
\beq
\partial^\mu \bar{\psi}\gamma_\mu \psi_L  = -im(\bar{\psi}_L\psi_R
-\bar{\psi}_R\psi_L) + \makebox{anomaly}
\eeq
\beq
\partial^\mu \bar{\psi}\gamma_\mu \psi_R  = -im(\bar{\psi}_R\psi_L
-\bar{\psi}_L\psi_R) + \makebox{anomaly}
\eeq
We thus need to subtract the vacuum to 2-gauge field 
matrix element
of the mass term, which is the operator $-im\bar{\psi}\gamma^5\psi$, 
to obtain the anomaly. The mass term yields 
a similar triangle diagram structure, and we define:
\bea
 M^5 & = & 
 (-1)(i)^2(i)^3\int' \int \frac{d^{4}\ell}{(2\pi)^{4}}
\frac{N_1+N_2}{D} \nonumber \\
N_1 & = & (-i)(-im)\Tr[\gamma^5(\slash{\ell}-\slash{q}+m)
(\slash{\epsilon}^L_c L+\slash{\epsilon}^R_c R) (\slash{\ell}+m)
(\slash{\epsilon}^L_b L+\slash{\epsilon}^R_b R) (\slash{\ell}+\slash{{k}}+m)]
\nonumber \\
N_2 & = & (+i)(-im)\Tr[\gamma^5(\slash{\ell}+\slash{k}-m)
(\slash{\epsilon}^L_b L+\slash{\epsilon}^R_b R)(\slash{\ell}-m)
(\slash{\epsilon}^L_c L+\slash{\epsilon}^R_c R)(\slash{\ell}-\slash{q}-m)]
\nonumber \\
D &  = &( \ell^2 +2\ell\cdot(zk-(1-y)q)+zk^2 +(1-y)q^2 -m^2 )^3
\eea
The result is (see \cite{hill2} for details):
\bea
M^5 & = & \bra{ 0} -im\bar{\psi}\gamma^5\psi \ket{b,c} = \frac{1}{24\pi^2} 
[2(E^{LL} + E^{RR})+ (E^{LR} + E^{RL})] \; , 
\eea
or, the operator correspondence:
\bea
im\bar{\psi}\gamma^5\psi  \rightarrow
-\frac{1}{48\pi^2} 
[F_L\tilde{F}_L + F_R\tilde{F}_R + F_L\tilde{F}_R] \; .
\eea
Forming the difference of the current
divergence with $-im\bar{\psi}\gamma^5\psi$ we have:
\bea
\label{div}
\partial_\mu\bar{\psi}\gamma_\mu\psi_L  
+  im(\bar{\psi}_L\psi_R
-\bar{\psi}_R\psi_L)
 & = &   \frac{1}{48\pi^2}(
F_L\tilde{F}_R+F_R\tilde{F}_R)
-\frac{1}{48\pi^2} 
[F_L\tilde{F}_L + F_R\tilde{F}_R + F_L\tilde{F}_R]
\nonumber \\
& = &   -\frac{1}{48\pi^2} 
F_L\tilde{F}_L \; .
\eea
Likewise:
\bea
\label{divR}
 \partial_\mu\bar{\psi}\gamma_\mu\psi_R  
+  im(\bar{\psi}_R\psi_L
-\bar{\psi}_L\psi_R)
  & = & - \frac{1}{48\pi^2}(
F_R\tilde{F}_L+F_L\tilde{F}_L)
+\frac{1}{48\pi^2} 
[F_L\tilde{F}_L + F_R\tilde{F}_R + F_L\tilde{F}_R]
\nonumber \\
& =  &  \frac{1}{48\pi^2} 
F_R\tilde{F}_R\; .
\eea

\newpage
\subsection{Summary}

\noindent
{\bf Pseudoscalar Mass Term}:
\bea
im\bar{\psi}\gamma^5\psi  \rightarrow
-\frac{1}{48\pi^2} 
[F_{L\mu\nu} \tilde{F}_L^{\mu\nu}+F_{R\mu\nu} \tilde{F}_R^{\mu\nu}
+F_{L\mu\nu} \tilde{F}_R^{\mu\nu}]
\eea

\noindent
{\bf Consistent Anomalies}:
\vskip 0.2in

\noindent
(1) Pure Massless Weyl Spinors ($p_i\cdot p_j >> m^2$):
\bea
\partial^\mu \bar{\psi}\gamma_\mu \psi_L  & = &
-\frac{1}{48\pi^2} F_{L\mu\nu} \tilde{F}_L^{\mu\nu}
\nonumber \\
\partial^\mu \bar{\psi}\gamma_\mu \psi_R  & = &
\frac{1}{48\pi^2} F_{R\mu\nu} \tilde{F}_R^{\mu\nu}
\eea

\noindent
(2) Heavy Massive Weyl Spinors ($p_i\cdot p_j << m^2$):
\bea
\partial^\mu \bar{\psi}\gamma_\mu \psi_L  +im(\bar{\psi}_L\psi_R
-\bar{\psi}_R\psi_L) &  = & -\frac{1}{48\pi^2} F_{L\mu\nu} \tilde{F}_L^{\mu\nu}
\nonumber \\
\partial^\mu \bar{\psi}_R\gamma_\mu \psi_R  + im(\bar{\psi}_R\psi_L
-\bar{\psi}_L\psi_R) & = & \frac{1}{48\pi^2} F_{R\mu\nu} \tilde{F}_R^{\mu\nu}
\eea

\noindent
(3) Heavy Massive Weyl Spinors ($p_i\cdot p_j << m^2$):
\bea
\partial^\mu \bar{\psi}\gamma_\mu \psi_L   &  = & 
\frac{1}{48\pi^2}(
F_{L\mu\nu} \tilde{F}_R^{\mu\nu} +
 F_{R\mu\nu} \tilde{F}_R^{\mu\nu} )
\nonumber \\
\partial^\mu \bar{\psi}\gamma_\mu \psi_R   
& = & -\frac{1}{48\pi^2} (
F_{L\mu\nu} \tilde{F}_R^{\mu\nu}
+F_{L\mu\nu} \tilde{F}_L^{\mu\nu})
\eea

\vskip 0.2in

\noindent
{\bf Consistent $L=V-A$ and $R=V+A$ Forms}:
\vskip 0.2in

\noindent
(1) Pure Massless Weyl Spinors ($p_i\cdot p_j >> m^2$):
\bea
\partial^\mu \bar{\psi}\gamma_\mu \psi  & = &
\frac{1}{12\pi^2} F_{V\mu\nu} \tilde{F}_A^{\mu\nu}
\nonumber \\
\partial^\mu \bar{\psi}\gamma_\mu\gamma^5 \psi  & = &
\frac{1}{24\pi^2}( F_{V\mu\nu} \tilde{F}_V^{\mu\nu} 
+ F_{A\mu\nu} \tilde{F}_A^{\mu\nu} )
\eea

\noindent
(2) Heavy Massive Weyl Spinors ($p_i\cdot p_j << m^2$):
\bea
\partial^\mu \bar{\psi}\gamma_\mu \psi &  = & \frac{1}{12\pi^2} F_{V\mu\nu} \tilde{F}_A^{\mu\nu}
\nonumber \\
\partial^\mu \bar{\psi}\gamma_\mu \gamma^5\psi  -2im\bar{\psi}\gamma^5\psi
& = & \frac{1}{24\pi^2}( F_{V\mu\nu} \tilde{F}_V^{\mu\nu} 
+ F_{A\mu\nu} \tilde{F}_A^{\mu\nu} )
\eea

\noindent
(3) Heavy Massive Weyl Spinors ($p_i\cdot p_j << m^2$):
\bea
\partial^\mu \bar{\psi}\gamma_\mu \psi   &  = & 
\frac{1}{12\pi^2} F_{V\mu\nu} \tilde{F}_A^{\mu\nu}
\nonumber \\
\partial^\mu \bar{\psi}\gamma_\mu \gamma^5 \psi   
& = & -\frac{1}{12\pi^2}( F_{V\mu\nu} \tilde{F}_V^{\mu\nu} )
\eea

\noindent
{\bf Covariant Forms}:
\vskip 0.2in

\noindent
Add a term to the lagrangian of
the form $(1/6\pi^2)\epsilon_{\mu\nu\rho\sigma}A^\mu V^\nu
\partial^\rho V^\sigma $. The currents are now modified to
$\tilde{J} = J +\delta J$ and $\tilde{J}^5 = J^5 + \delta J^5$
as described in the text.

\noindent
(1) Pure Massless Weyl Spinors ($p_i\cdot p_j >> m^2$):
\bea
\partial^\mu \tilde{J}_\mu  & = & 0
\nonumber \\
\partial^\mu \tilde{J}^5_\mu  & = &
\frac{1}{8\pi^2}( F_{V\mu\nu} \tilde{F}_V^{\mu\nu} 
+ \frac{1}{3}F_{A\mu\nu} \tilde{F}_A^{\mu\nu} )
\eea

\noindent
(2) Heavy Massive Weyl Spinors ($p_i\cdot p_j << m^2$):
\bea
\partial^\mu \tilde{J}_\mu &  = & 0
\nonumber \\
\partial^\mu \tilde{J}^5_\mu   - 2im\bar{\psi}\gamma^5\psi
& = & \frac{1}{8\pi^2}( F_{V\mu\nu} \tilde{F}_V^{\mu\nu} 
+\frac{1}{3} F_{A\mu\nu} \tilde{F}_A^{\mu\nu} )
\eea

\noindent
(3) Heavy Massive Weyl Spinors ($p_i\cdot p_j << m^2$):
\bea
\partial^\mu \tilde{J}_\mu   &  = & 0
\nonumber \\
\partial^\mu \tilde{J}^5_\mu  
& = & 0
\eea
The latter case is completely summarized by the fact
that, for KK-modes, 
the three-gauge boson amplitude is described by the operator:
\beq
\label{triop2}
{\cal{O}}_3 =  -\frac{1}{12\pi^2}\epsilon^{\mu\nu\rho\sigma} 
\sum_{nmk}a_{nmk}B^n_\mu B^m_\nu\partial_\rho B_\sigma^k 
\eeq
where:
\beq
a_{nmk} = \half [1 - (-1)^{n+m+k}](-1)^{m+k}
\eeq
This operator is equivalent to 
$(-1/6\pi^2)\epsilon_{\mu\nu\rho\sigma}A^\mu V^\nu
\partial^\rho V^\sigma $
when we truncate on the first two KK-modes,
and identify $B^0 =  V$ and $B^1=A$.
Adding the
$(1/6\pi^2)\epsilon_{\mu\nu\rho\sigma}A^\mu V^\nu
\partial^\rho V^\sigma $ term cancels this quantity,
completely cancels the triangle diagrams,
and the resulting currents then have vanishing divergences.

\newpage 
\vskip .2in
\noindent
{\bf Acknowledgments}
\vskip .1in
We especially thank 
Bill Bardeen and Bogdan Dobrescu for numerous
helpful discussions. We also wish to thank Tao Han and
Sherwin Love for helpful discussions in the early phase of this work.
This work is supported in part by
the US Department of Energy, High Energy Physics Division,
Contract W-31-109-ENG-38, and 
grant DE-AC02-76CHO3000.

\end{document}